\documentclass[aps,prb,twocolumn]{revtex4}%
\usepackage{amsfonts}
\usepackage{amsmath}
\usepackage{amssymb}
\usepackage{psfrag}
\usepackage{graphicx}
\usepackage{dsfont}%

\newlength{\graphwid}
\def\showgraph#1#2{
\settowidth{\graphwid}{\includegraphics[#1,clip=true]{#2}}
\parbox[c]{\graphwid}{\includegraphics[#1,clip=true]{#2}}}
\ifx\pdfoutput\relax\let\pdfoutput=\undefined\fi
\newcount\msipdfoutput
\ifx\pdfoutput\undefined\else
\ifcase\pdfoutput\else
\ifx\paperwidth\undefined\else
\ifdim\paperheight=0pt\relax\else\pdfpageheight\paperheight\fi
\ifdim\paperwidth=0pt\relax\else\pdfpagewidth\paperwidth\fi
\fi\fi\fi
\begin{document}
\preprint{ }
\title[ ]{$\mathbf{J_{1}}$-$\mathbf{J_{2}}$ frustrated two-dimensional Heisenberg model: Random phase approximation and functional renormalization group}
\author{Johannes Reuther and Peter W\"{o}lfle}
\affiliation{Institut f\"{u}r Theorie der Kondensierten Materie, Karlsruhe Institute
of Technology, D-76128 Karlsruhe, Germany}
\author{}
\affiliation{}

\begin{abstract}
We study the ground state properties of the two-dimensional spin-1/2 $J_{1}
$-$J_{2}$-Heisenberg model on a square lattice, within diagrammatic
approximations using an auxiliary fermion formulation with exact projection.
In a first approximation we assume a phenomenological width of the
pseudofermion spectral function to calculate the magnetization,
susceptibilities and the spin correlation length within RPA, demonstrating the
appearance of a paramagnetic phase between the N\'{e}el ordered and Collinear
ordered phases, at sufficiently large pseudo fermion damping. Secondly we use
a Functional Renormalization Group formulation. We find that the conventional
truncation scheme omitting three-particle and higher order vertices is not
sufficient. We therefore include self-energy renormalizations in the 
single-scale propagator as recently proposed by Katanin, to preserve Ward identities in a
better way. We find N\'{e}el order at $g=J_{2}/J_{1}\lesssim g_{c1}
\approx0.4\ldots0.45$ and Collinear order at $g\gtrsim g_{c2}\approx
0.66\ldots0.68$, which is in good agreement with results obtained by numerical
studies. In the intervening quantum paramagnetic phase we find enhanced
columnar dimer and plaquette fluctuations of equal strength.

\end{abstract}



\maketitle

\section{Introduction}{\label{sec1}} 

It has been known for a long time that quantum antiferromagnets, i.e. spin-1/2
systems coupled by Heisenberg exchange interaction, are strongly affected by
quantum fluctuations at low temperatures. Thermal fluctuations are important
as well, especially since they suppress long-range-order (LRO) in two
dimensions at any finite temperature, but their role is relatively well
understood. By contrast, quantum fluctuations operate in a much more complex
way: they may suppress LRO, but may at the same time lead to novel ground
states known under the labels \textquotedblleft spin liquids, valence-bond
solids\textquotedblright. The first such state proposed in the literature is
Anderson's RVB-(\textquotedblleft resonating-valence-bond\textquotedblright%
)-state\cite{a1}. In the context of cuprate superconductors, viewed as hole-doped
Mott insulators, RVB-states have been proposed by Anderson\cite{rvb1}  to form the
fundamental basis on which the theory of high-Tc superconductivity should be
built. Although the idea has been considered by many authors since then, there
is no conclusive answer to the question of the role of a spin liquid state for
HTSC. These studies have raised the question, however, under which conditions
quantum fluctuations are strong enough to destroy long range order. In
general, spin liquid type states may be expected to be stabilized by any type
of quantum fluctuations. For spin systems it has been proposed that
frustration either by competing spin interactions or due to special geometric
arrangements may lead to a spin liquid state. In particular, by tuning the
interactions or the lattice anisotropy a quantum phase transition from a state
with long-range order into a spin liquid state may take place. Generally
speaking it has proven to be, may be unexpectedly, hard to destroy long range
order by quantum fluctuations.

The simplest theoretical model of such a system is the Quantum Heisenberg
Antiferromagnet (QHAF) with nearest neighbor interaction for spins
${\frac12}$
on a two-dimensional square lattice. Its ground state is known to be the N\'{e}el
state with staggered magnetization reduced by quantum fluctuations\cite{spinwave}. At any
finite temperature the magnetic order is destroyed by thermal fluctuations,
but the correlation length is found to increase exponentially with decreasing
temperature\cite{a4,a5}. This physics has also been obtained from the Quantum
Nonlinear Sigma Model (QNL$\sigma$M) in the renormalized classical regime\cite{a6}.

A simple model with competing interactions is the $J_{1}$-$J_{2}$ model\cite{mod2},
featuring an additional antiferromagnetic next-nearest neighbor interaction
$J_{2}$ in addition to the nearest neighbor coupling $J_{1}$ . This model has
attracted attention as a simplified model\cite{a8} for the effect of doping in the
cuprate superconductors: when a small concentration of holes is doped into the
CuO-planes, the long-range AF order of the undoped system is rapidly 
destroyed\cite{a9,a12}, giving way to a non-magnetic \textquotedblleft
pseudo-gap\textquotedblright\ state and to superconductivity.

Recently, this model has also found use for certain Vanadate compounds\cite{exp1,exp2}, for which the magnetic interactions can be modeled by the
$J_{1}$-$J_{2}$ Hamiltonian of weakly coupled planes.

Even more recently the $J_{1}$-$J_{2}$ model has been invoked to account for the
weakened AF long range order in the iron pnictides\cite{a15,a16,a17}. The universally
observed linear temperature dependence of the magnetic susceptibility of these
compounds has also been addressed in the framework of the $J_{1}$-$J_{2}$ model\cite{a18}.

If the spins of the model are considered as classical ($S\rightarrow\infty$),
there is an abrupt transition from N\'{e}el order to the Collinear configuration
for sufficiently strong frustration, $J_{2}/J_{1}=1/2$. In mean field
approximation a first order transition from the N\'{e}el to the Collinear state is
found. However, this is changed by quantum fluctuations. Early on it has been
found\cite{mod2,mod11,mod6} that a non-magnetic phase exists in the region $0.4\lesssim J_{1}%
/J_{2}\lesssim 0.65$, between the two ordered states. The nature of this intermediate
state is what we would like to unravel. Most recent mainly numerical work 
(see Refs. \onlinecite{sirker,darradi,isaev} and references therein) on the $J_{1}$-$J_{2}$ model indicates that
it may be a valence-bond solid\cite{mod5} (VBS), rather than a homogeneous spin
liquid\cite{dimer4}. In the VBS the spins in the plane form pairwise singlets, which
are spontaneously dimerized in a, e.g. columnar pattern and therefore break
the lattice translational symmetry. It has also been proposed that the
dimerization takes place on units of 2$\times$2 plaquettes\cite{mod13}. Evidence for a VBS
has also been found in studies\cite{a27,a28} of a model of coupled spin chains\cite{a29},
when the results are extrapolated to the isotropic $J_{1}$-$J_{2}$ model in the
plane. Concerning the nature of the quantum phase transitions recent studies
indicate that the transition from the paramagnetic phase to the Collinear
configuration is of first order\cite{sirker,dimer2,darradi,isaev}. On the other hand the
properties of the transition from the N\'{e}el phase to the paramagnetic phase are
still highly controversial. Recent studies point to either a first order\cite{a27,sirker} or 
a second order transition\cite{dimer2,dimer3}. The latter scenario gives rise
to the question of how two differently ordered phases may be connected by a
continuous phase transition\cite{deconf1}.

In this paper we develop new analytical methods for calculating the ground
states and the excitation spectra of spin models with competing interactions,
such as the model discussed above, on the basis of infinite resummations of
perturbation theory in the couplings $J_{1}$, $J_{2}$. To this end we use a
representation of the spin operators in terms of pseudofermions\cite{abrik}. One
motivation for using a fermionic representation rather than a bosonic
representation is the available experience in describing spin liquids or
dimerized spin-singlet states with fermions, mainly within large-N and
Mean-Field approaches (see e.g. Refs. \onlinecite{rvb2,a35,a36,a37}). On the other hand,
pseudofermion representations have hardly been used to study magnetic ordering
phenomena\cite{a38}. Although a large body of results of numerical studies of these
models is available, analytical approaches starting from a microscopic
Hamiltonian are rare. We use a newly developed implementation of the
Functional Renormalization Group (FRG) method\cite{katanin,kat1} applied to interacting
quantum spin models. In this we are aided by the experience we have previously
gained with the nearest neighbor Heisenberg model\cite{jan1,jan2}. Auxiliary particle
representations of spin operators are sometimes viewed with suspicion, as they
are conceived to be fraught with uncontrolled approximations regarding the
projection unto the physical sector of the Hilbert space necessary in those
spin representations. Here we are using an exact method of projection onto the
physical part of Hilbert space that works even on the lattice (see below). 

The paper is organized as follows: Sec. \ref{sec2} introduces the model, the
auxiliary-fermion representation and the projection schemes in detail. Simple
Mean-Field approximations are discussed in Sec. \ref{sec3} where we demonstrate that these approaches are not able to capture frustration effects but rather reproduce classical results. To this end in Sec. \ref{sec4} we introduce a phenomenological pseudo-particle lifetime that mimics quantum fluctuations. The results on the magnetization, susceptibilities and spatial spin correlations show that in a certain parameter range for this lifetime, the correct phase diagram is obtained. The main part of the paper, given by Sec. \ref{sec5} is devoted to FRG. This method enables us to calculate the auxiliary particle damping rather than treating it as an input of the approximation. In Sec. \ref{subsec51} we first point out that the often applied static FRG scheme does not lead to a finite flow of the damping. Therefore in Sec. \ref{subsec52}, in the framework of the standard truncation of the FRG equations we include the full dynamics. It turns out that within the latter (one-loop) approximation the strength of quantum fluctuations in the highly frustrated region is still underestimated, and a regime without N\'{e}el order or Collinear order is not found. We trace this deficiency of the one-loop approximation to the neglect of higher order contributions. Another way of saying this is that the Ward identities resulting from spin conservation are badly violated in the one-loop scheme, such that not even the RPA approximation is reproduced in that approximation. As shown by Katanin\cite{katanin} the latter problem may be remedied by using a dressed single scale propagator, thus including three-particle correlations with non-overlapping loops. As shown in Sec. \ref{subsec53}, using the Katanin truncation scheme we find a phase diagram in excellent agreement with results from numerical methods. In order to investigate the properties of the non-magnetic phase, correlation functions for columnar dimer and plaquette order are calculated in Sec. \ref{subsec54}. We find that correlations for both kinds of dimerizations are clearly enhanced. Finally, the paper closes with a summary in Sec. \ref{sec6}.

It is worth mentioning that although in this first presentation of our work using the newly developed FRG method we concentrated on demonstrating that the method is capable of giving results in agreement with results obtained mainly by purely numerical means, it should be clear that the method holds in fact considerable promise for future applications. First of all, it allows to treat thermodynamic, in contrast to finite size systems. Second, it is ready to calculate dynamical properties (at least on the imaginary frequency axis). Thirdly, it is easily generalized to finite temperature (work in progress). Further, it allows in principle to address questions of critical behavior near a quantum critical point.

\section{The model}{\label{sec2}}

The effects of frustration in quantum spin models have been intensely studied
in recent years. These models offer the possibility to investigate quantum phase
transitions\cite{sachdev} between magnetically ordered and disordered phases.
Especially in the context of deconfined criticality in two dimensional spin
systems\cite{deconf1,deconf2}, quantum phase transitions are the object of
renewed interest. A standard model capturing these phenomena is the spin-$1/2$
Heisenberg model on a square lattice with an antiferromagnetic nearest
neighbor coupling $J_{1}>0$ and a frustrating next-nearest neighbor coupling
$J_{2}\geq0$, see Refs.
\onlinecite{mod2,mod3,mod5,mod6,mod11,mod12,mod13,mod15,mod16,dimer1,mod19,dimer2,mod21,dimer3,dimer4,sirker,mod28,mods7,darradi,isaev},
\begin{equation}
H=J_{1}\sum_{nn}\mathbf{S}_{i}\cdot\mathbf{S}_{j}+J_{2}\sum_{nnn}%
\mathbf{S}_{i}\cdot\mathbf{S}_{j}\quad.\label{1}%
\end{equation}

As far as the ground state of the model is concerned, two limiting cases are
well understood: For $J_{2}=0$ the system is in a N\'{e}el-ordered phase with
a magnetization of $\sim60\%$ of the saturation magnetization. In the limit
$g=\frac{J_{2}}{J_{1}}\rightarrow\infty$ the model reduces to two decoupled
square lattices. N\'{e}el order on each of these lattices gives rise to the
so-called Collinear long-range order with magnetic wave vectors $Q=(\pi,0)$ or
$Q=(0,\pi)$. These two degenerate ground states correspond to a parallel
alignment of the spins along the $y$-axes and a antiparallel alignment of
neighboring spins along the $x$-axes in the first case and vice versa in the
second case. Increasing $J_{2}$ in the first limit or $J_{1}$ in the second
limit leads in both cases to frustration and to a decrease of the respective
order parameter, possibly all the way to zero. Indeed, the existence of a
non-magnetic phase, indicated by numerical studies, approximately in the
parameter region $g\approx0.4\ldots0.65$ is widely
accepted\cite{mod6,mod12,mod16,dimer2,darradi,isaev}. However, the nature of
the magnetically disordered phase as well as the order of the quantum phase
transitions is not known with certainty so far. Candidates for this phase are
a spin liquid \cite{mod2,mod19,mod21} and a valence-bond solid state (VBS).
For the latter, different types of order have been proposed, e.g., columnar
dimer \cite{mod16,dimer2,sirker,mods7} and plaquette
\cite{mod13,dimer1,dimer2,isaev} order. Several studies give evidence that the
transition from the non-magnetic phase to the Collinear phase is of first
order \cite{mod16,dimer2,sirker,darradi,isaev}. We also mention that in the
classical large spin-limit no magnetically disordered phase exists. Instead
there is a direct first order transition between the N\'{e}el phase and the
Collinear phase at $g=1/2$. In this limit the respective
magnetizations reach the saturation value.

In the past the model has been studied with a variety of methods. Examples
are: analytical approaches based on field-theory methods\cite{mod5,dimer3} or
spin-wave theory\cite{mod2,mod11}; numerical approaches such as exact
diagonalization\cite{mod6,mod12,dimer1,dimer4,mod28}, coupled cluster
method\cite{mod15,darradi}, series expansion
methods\cite{mod16,dimer2,dimer3,sirker,mods7} and Quantum Monte Carlo
method\cite{dimer1,mod21} .

In this paper we address the ground state properties in a rather different
way. In order to allow for application of Feynman-diagram
techniques\cite{negele,rickayzen} we reformulate the problem in a fermionic
language by introducing auxiliary fermions \cite{abrik,jan1,jan2}. We
represent the spin operators in terms of auxiliary fermions $f_{i\alpha}$,
\begin{equation}
S_{i}^{\mu}=\frac{1}{2}\sum_{\alpha\beta}f_{i\alpha}^{\dagger}\sigma
_{\alpha\beta}^{\mu}f_{i\beta}\quad.\label{3}
\end{equation}
Here $\sigma^{\mu}$ ($\mu=x,y,z$) are Pauli matrices, $\alpha,\beta
=\uparrow,\downarrow$ are spin indices and $i$ is the site index. We use units
with $\hbar=k_{\mathrm{B}}\equiv1$. By construction, the representation
(\ref{3}) satisfies the correct commutation relations. However, the
Hilbert space for a single site $i$ is now spanned by four states, of which
two, representing an empty and a doubly occupied site are unphysical. The
projection to the physical sector of Hilbert space is given by the
auxiliary-particle constraint
\begin{equation}
Q_{i}=\sum_{\alpha}f_{i\alpha}^{\dagger}f_{i\alpha}=1
\end{equation}

We present two different projection schemes to account for this constraint.

A convenient approximate approach is to replace the constraint $Q_{i}=1$ by
its thermodynamic average, $\langle Q_{i}\rangle=1$ . For a
translationinvariant state the latter conditions are identical at each site,
such that only a single condition remains. Since the constraint amounts to
removing two of the four states per site, it is on average equivalent to
half-filling of the system, which in case of particle-hole symmetry is
effected by applying a chemical potential $\mu=0$ \ to the pseudofermion system.

A different approach allowing for an exact treatment of the constraint even
for lattice systems has been proposed by Popov and Fedotov\cite{popov}. It
amounts to applying a homogeneous, \textit{imaginary-valued} chemical
potential $\mu^{\text{ppv}}=-\frac{i\pi T}{2}$ , where $T$ is the temperature.
Thus, within this scheme, the Hamiltonian $H$ is replaced by
\begin{equation}
H\longrightarrow H^{\text{ppv}}=H-\mu^{\text{ppv}}\sum_{i}Q_{i}\quad.
\end{equation}
Note that $H$ denotes the Hamiltonian (\ref{1}) using the representation of
spin operators (\ref{3}). Given a physical operator $\mathcal{O}$ (i.e., an
arbitrary sum or product of spin operators) it can be shown\cite{jan2} that
the expectation value $\langle\mathcal{O}\rangle^{\text{ppv}}$, calculated
with $H^{\text{ppv}}$ and the \textit{entire} Hilbert space, is identical to
the physical expectation value $\langle\mathcal{O}\rangle$, where the average
is performed with the original Hamiltonian $H$ . The projection works by
virtue of a mutual cancellation of the unphysical contributions of the sectors
$Q_{i}=0$ and $Q_{i}=2$, at each site. It should be emphasized that
although the Hamiltonian $H^{\text{ppv}}$ is no longer hermitian, the quantity
$\langle\mathcal{O}\rangle^{\text{ppv}}$ comes out real-valued. If on the
other hand $\mathcal{O}$ is unphysical in the sense that it is non-zero in the
unphysical sector, e.g., the operator $\mathcal{O}=Q_{i}$, the expectation
value $\langle Q_{i}\rangle^{\text{ppv}}$ is meaningless and one has $\langle
Q_{i}\rangle\neq\langle Q_{i}\rangle^{\text{ppv}}$.

This approach is applicable to spin models\cite{jan2,useppv1,useppv2} but
unfortunately it can not be extended to cases away from half filling. Although
$\mu^{\text{ppv}}$ vanishes in the limit $T\rightarrow0$, in principle the
exact projection with $\mu=\mu^{\text{ppv}}$ and the average projection with
$\mu=0$ are not equivalent at $T=0$. This is due to the fact that the
computation of an average $\langle\dots\rangle^{\text{ppv}}$ does not
necessarily commute with the limit $T\rightarrow0$. Nevertheless it can be
expected that in the model considered here both projection schemes are
identical at $T=0$. This can be understood with the following argument:
Starting from the physical (\textquotedblleft true\textquotedblleft) ground
state, a fluctuation of the pseudofermion number results in two sites with
unphysical occupation numbers, one with no and one with two fermions. Since
these sites carry spin zero the sector of the Hamiltonian with that occupation
is identical to the physical Hamiltonian where the two sites are effectively
\textit{missing}. Thus a fluctuation from the ground state into this sector
costs the binding energy of the two sites which is of the order of the
exchange coupling, even in the case of strong
frustration\cite{mod13,mod16,sirker,darradi,isaev}. Consequently, at $T=0$
pseudofermion-number fluctuations are not allowed and it is sufficient to use the simpler
average projection with $\mu=0$. In most calculations we restrict ourself to
this method. However, we again emphasize that at $T>0$ both schemes differ.

In the following we will formulate approximations in terms of resummed
perturbation theory in the exchange couplings $J_{1}$, $J_{2}$ . The basic building
blocks are the four-fermion interactions and the bare fermion Green's function
in real space
\begin{equation}
G_{ij,\alpha\beta}^{0}(i\omega)=\frac{1}{i\omega+\mu}\delta_{ij}\delta
_{\alpha\beta}\;;\;\mu=-\frac{i\pi T}{2}\;\mathrm{or}\;\mu=0\;.\label{7}
\end{equation}
$\omega=(2n+1)\pi T$ are the fermionic Matsubara-frequencies. Note that in
diagrammatic expansions the Green's functions remain strictly local, i.e.,
$G_{ij,\alpha\beta}=\delta_{ij}G_{i,\alpha\beta}$. The momentum dependence in
correlators like the susceptibility is generated by the non-local exchange couplings.

We begin the calculations with a simple Mean-Field approach. It should be
stressed that in our model a small parameter is absent. Accordingly, a
controlled summation of diagrams is a difficult task. For this reason we
extend the Mean-Field approach and set up a phenomenological theory which
explores the consequences of certain assumptions on the width of the auxiliary
fermion spectral-function and which gives qualitatively correct results.

Furthermore, the feasibility of diagrammatic approximations allows the
application of the Functional Renormalization Group method
(FRG)\cite{frg1,frg2,frg3,frg4}. This scheme generates an exact, infinite
hierarchy of coupled differential equations for the one-particle irreducible
m-particle vertex functions by introducing an infrared cutoff. In order to be
able to solve these equations numerically one truncates the hierarchy of
equations. The truncation is expected to give good results for not too strong
interaction. It will turn out that the truncation procedure is a non-trivial
problem for the model considered here. Effectively FRG sums up infinite
classes of diagrams. This is a crucial property in the present problem for
which a small parameter does not exist.

So far, FRG has been applied to low dimensional, interacting fermion systems,
e.g., the two dimensional Hubbard model\cite{frg2,hubbard}, the single
impurity Anderson model\cite{anderson} and the Luttinger liquid with
impurities\cite{luttinger}, but a pure spin model has not been tackled with FRG.

\section{Mean-Field theory}{\label{sec3}} 

\begin{figure}[t]
\begin{center}
\settowidth{\graphwid}{\includegraphics[scale=0.7,clip=true]{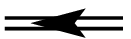}}
\parbox[c]{\graphwid}{\includegraphics[scale=0.7,clip=true]{grenren.eps}}\;\;=\negthickspace
\settowidth{\graphwid}{\includegraphics[scale=0.7,clip=true]{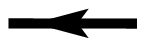}}
\parbox[c]{\graphwid}{\includegraphics[scale=0.7,clip=true]{gren.eps}}\;+\;\raisebox{-0.04cm}{\includegraphics[scale=0.7]{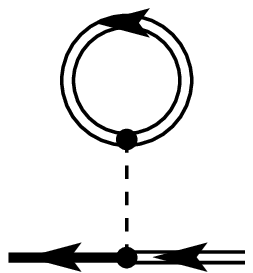}}
\end{center}
\caption{Diagrammatic representation of the Hartree approximation. The full
line is the bare Green's function $G^{0}$, Eq. (\ref{7}), the double-stroke
line the self consistent one. The dashed line represents the interaction
$J_{1}$ or $J_{2}$ and the dots are Pauli matrices$\times$1/2.}%
\label{fig1}
\end{figure}

The most elementary approximation for a spin model is the
Mean-Field theory. In our fermionic description it corresponds to the Hartree
approximation shown in Fig. \ref{fig1}. The closed loop of the renormalized
propagator acts as the self-consistent Mean-Field.

Note that the Fock term is exactly zero, since the non-local exchange coupling connects two points of the same fermion line. Dropping the requirement of exact projection, one may allow for fermion hopping and make a Mean-Field ansatz with nonlocal propagators. The corresponding symmetry-broken phase is the so called
resonating valence bond (RVB)\cite{rvb1} or the flux phase\cite{rvb2,rvb3}. We
will not consider Mean-Field amplitudes violating the auxiliary particle
constraint in this paper.

By contrast, the magnetic order parameter $\langle\mathbf{S}_{i}\rangle
=\frac{1}{2}\sum_{\alpha\beta}\langle f_{i\alpha}^{\dagger}%
\mbox{\boldmath$\sigma$}_{\alpha\beta}f_{i\beta}\rangle$ that appears in the
Hartree approximation is a physical quantity. In the following calculation we
also consider finite temperatures. Dyson's equation in Fig. \ref{fig1} reads
\begin{equation}
\bar{G}_{i}(i\omega)=[(i\omega+\mu)\mathds{1}-\bar{\Sigma}_{i}(i\omega
)]^{-1}\quad,\label{8}%
\end{equation}
where $\bar{G}$, $\bar{\Sigma}$ and $\mathds{1}$ are matrices in spin space.
The self-energy is coupled back to the renormalized Green's function by
\begin{equation}
\bar{\Sigma}_{i}(i\omega)=\frac{1}{4}\sum_{j}J_{ij}\sum_{\mu=1}^{3}\sigma
^{\mu}\frac{1}{\beta}\sum_{i\omega'}\mathrm{Tr}[\sigma^{\mu}\bar{G}_{j}%
(i\omega')]e^{i\omega'\delta}\quad.\label{9}%
\end{equation}
The couplings are written in the form $J_{ij}$, which is $J_{1}$ if $i,j$ are
nearest neighbors, and $J_{2}$ if $i,j$ are next nearest neighbors. The factor
$e^{i\omega\delta}$, with an infinitesimal $\delta>0$, is needed for the
convergence of the Matsubara sum. If we assume magnetism along the
$z$-direction, the self-energy has the form
\begin{equation}
\bar{\Sigma}_{i}(i\omega)=\sigma^{z}m_{i}\quad.\label{10}%
\end{equation}
To describe N\'{e}el- and Collinear order we split the lattice up into two
sublattices $A$ and $B$. In case of N\'{e}el order $A$ and $B$ form a
staggered pattern while for Collinear order they form rows (or equivalently
columns). Furthermore we require
\begin{equation}
m\equiv m_{i\in A}=-m_{i\in B}\quad.\label{11}%
\end{equation}
Inserting Eq. (\ref{8}) into Eq. (\ref{9}) and using Eq. (\ref{10}) one
obtains
\begin{equation}
m_{i}=\frac{1}{4}\sum_{j}J_{ij}\frac{1}{\beta}\sum_{i\omega}\sum_{\zeta=\pm
1}\frac{\zeta}{i\omega+\mu-\zeta m_{j}}e^{i\omega\delta}\quad.\label{12}%
\end{equation}
Using $\frac{1}{\beta}\sum_{i\omega}\frac{e^{i\omega\delta}}{i\omega-z}=f(z)$
and $f(z-\mu^{\text{ppv}})=\frac{1}{ie^{\beta z}+1}$ ($f$ is the Fermi
function) one finds the following self-consistent equations for $m$ for both
types of order and both projection schemes,
\begin{subequations}
\begin{align}
&  \hspace*{-10pt}\mbox{N\'{e}el-order: }m=%
\begin{cases}
(J_{1}-J_{2})\tanh(\frac{m\beta}{2}) & \mbox{for }\mu=0\\
(J_{1}-J_{2})\tanh(m\beta) & \mbox{for }\mu=\mu^{\mathrm{ppv}}%
\end{cases}
\;,\nonumber\\ \label{13a}
& \\
&  \hspace*{-10pt}\mbox{Collinear-order: }m=%
\begin{cases}
J_{2}\tanh(\frac{m\beta}{2}) & \mbox{for }\mu=0\\
J_{2}\tanh(m\beta) & \mbox{for }\mu=\mu^{\mathrm{ppv}}%
\end{cases}
\;.\nonumber\\ \label{13b}
&
\end{align}
The spin polarization or, in short, magnetization $M_{i}$ is given by
\end{subequations}
\begin{equation}
M_{i}=\langle S_{i}^{z}\rangle=\frac{1}{2}\frac{1}{\beta}\sum_{i\omega
}\mathrm{Tr}[\sigma^{z}\bar{G}_{i}(i\omega)]e^{i\omega\delta}\quad.\label{14}%
\end{equation}
From the comparison of Eq. (\ref{9}) and Eq. (\ref{14}) and using $M_{i\in
A}=-M_{i\in B}$ one finds a relation between $m_{i}$ and $M_{i}$, \
\begin{equation}
m_{i}\hspace*{-1pt}=\hspace*{-1pt}\frac{1}{2}\sum_{j}J_{ij}M_{j}
\hspace*{-1pt}=\hspace*{-1pt}%
\begin{cases}
2M_{i}(J_{2}-J_{1}) & \hspace*{-3.2pt}\mbox{for N\'{e}el-order}\\
-2M_{i}J_{2} & \hspace*{-3.2pt}\mbox{for Collinear order}
\end{cases}
.\label{15}%
\end{equation}
From Eqs. (\ref{13a}) and (\ref{13b}) the critical temperatures $T_{\mathrm{c}%
}^{\mathrm{N\acute{e}el}}$ and $T_{\mathrm{c}}^{\mathrm{Col}}$ can be
determined. The instability with the larger transition temperature controls
the type of order at a given $g=\frac{J_{2}}{J_{1}}$. This leads to
\begin{subequations}
\label{16b}%
\begin{align}
&  \hspace*{-8pt}0\leq g\leq\frac{1}{2}\;:\;T_{\mathrm{c}}=T_{\mathrm{c}%
}^{\mathrm{N\acute{e}el}}=%
\begin{cases}
\frac{J_{1}}{2}(1-g) & \mbox{for }\mu=0\\
J_{1}(1-g) & \mbox{for }\mu=\mu^{\mathrm{ppv}}%
\end{cases}
\;,\nonumber\\
& \\
&  \hspace*{-8pt}g\geq\frac{1}{2}\;:\;T_{\mathrm{c}}=T_{\mathrm{c}%
}^{\mathrm{Col}}=%
\begin{cases}
\frac{J_{1}}{2}g & \mbox{for }\mu=0\\
J_{1}g & \mbox{for }\mu=\mu^{\mathrm{ppv}}%
\end{cases}
\;.\nonumber\\
&
\end{align}

Apparently, within this approximation, no non-magnetic phase is found at $T=0$
. Instead there is a first order transition from N\'{e}el- to Collinear order
at $g=\frac{1}{2}$. The magnetization $M=|M_{i}|$, which can be obtained from
Eqs. (\ref{13a}), (\ref{13b}) and (\ref{15}), reaches the saturation value
$M=\frac{1}{2}$ at $T=0$, and the classical large spin behavior is reproduced.
These properties hold for both projection schemes. However, this is no longer
the case for $T>0$. The contribution of unphysical states with $S=0$ leads to
a reduction of the magnetization in the average projection scheme. Also the
critical temperatures come out a factor of two smaller in the average
projection scheme. The self-consistent equations for $\mu=\mu^{\mathrm{ppv}}$
are identical to those obtained within the conventional Mean-Field theory in
terms of spin operators, confirming that the cancellation of the unphysical
states works correctly in this approximation.

In summary, the simple Mean-Field theory leads to a N\'{e}el phase at
$g<\frac{1}{2}$ and a Collinear-ordered phase $g>\frac{1}{2}$ but is not
sufficient to describe the effect of frustration in destroying magnetic order
in the regime $g\approx\frac{1}{2}$.

\section{Finite pseudofermion lifetime}{\label{sec4}} 

In the Mean-Field approximation the effect of \ fermion scattering in
generating a finite lifetime is not taken into account. We now briefly discuss
a phenomenological framework for the ground state, which introduces the
lifetime $\tau$ as a phenomenological parameter. To this end \ we model the
retarded Green's function by,
\end{subequations}
\begin{equation}
G^{\mathrm{R}}(\omega)=\frac{1}{\omega+i\gamma}\;\;,\;\;\Sigma=-i\gamma
\;\;\mathrm{with}\;\;\gamma=\frac{1}{\tau}\quad.\label{18}%
\end{equation}
The spectral function $\rho(\omega)$ acquires a finite width $\gamma$,
\begin{equation}
\rho(\omega)=-\frac{1}{\pi}\mathrm{Im}\;G^{\mathrm{R}}(\omega)=\frac{\gamma}{\pi}%
\frac{1}{\omega^{2}+\gamma^{2}}\quad.\label{19}%
\end{equation}
An analytic continuation of the self-energy $\Sigma$ to the upper complex
half-plane provides
\begin{equation}
\Sigma(z)=-i\gamma\;\;\mathrm{for}\;\;\mathrm{Im}\;z>0\quad.\label{20}%
\end{equation}
Since $\rho(\omega)$ is an even function it follows immediately from the
spectral representation $G(i\omega)=\int_{-\infty}^{\infty}\frac{\rho
(\epsilon)}{i\omega-\epsilon}\mathrm{d}\epsilon$ that $G(z)$ and $\Sigma(z)$
are odd functions with vanishing real parts along the complex Matsubara-axes.
\begin{figure}[t]
\begin{center}
\includegraphics*[scale=0.8]{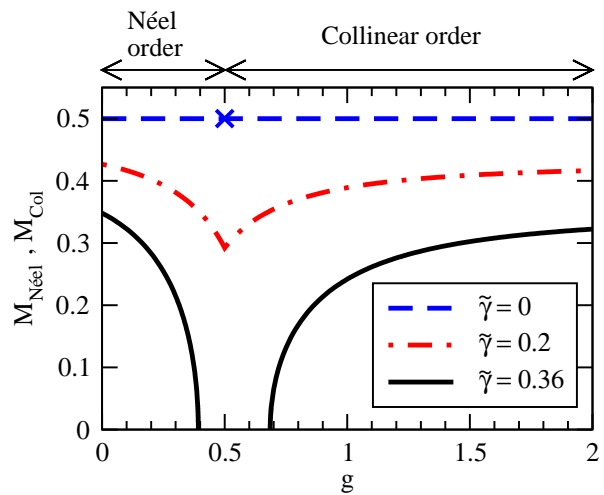}
\end{center}
\caption{Magnetizations $M_{\text{N\'{e}el}}$ and $M_{\text{Col}}$ versus $g$
within the phenomenological theory for different damping parameters
$\tilde{\gamma}$ (color online).}%
\label{fig2}%
\end{figure}
Thus we obtain
\begin{equation}
G(i\omega)=\frac{1}{i\omega+i\gamma\:\mathrm{sgn}(\omega)}\quad.\label{21}%
\end{equation}
To proceed, we need to specify the $g$-dependence of the damping parameter
$\gamma$. For $J_{2}=0$ we put $\gamma$ in the form $\gamma=\tilde{\gamma
}J_{1}$ where $\tilde{\gamma}$ is a dimensionless parameter. A similar
situation is encountered for $J_{1}\rightarrow0$, $J_{2}>0$, where the system
is split up into two square lattices, each only with nearest-neighbor
couplings $J_{2}$. Therefore in this limit the relation $\gamma=\tilde{\gamma
}J_{2}$ holds. To interpolate between both limiting cases we assume
\begin{equation}
\gamma(J_{1},J_{2})=\tilde{\gamma}J_{1}\sqrt{1+g^{2}}\quad.\label{22}%
\end{equation}

\subsection{Hartree approximation}{\label{subsec41}} 

Replacing the bare Green's function by Eq. (\ref{21}) we
now calculate the ground state magnetization within the Hartree approximation
of Sec. \ref{sec3}. In the limit $T\rightarrow0$, using $\frac{1}{\beta}%
\sum_{i\omega}\rightarrow\frac{1}{2\pi}\int\mathrm{d}\omega$ , the new Mean-
Field equation is given by
\begin{equation}
m_{i}=\frac{1}{4}\sum_{j}J_{ij}\frac{1}{2\pi}\int_{-\infty}^{\infty
}\mathrm{d}\omega\sum_{\zeta=\pm1}\frac{\zeta}{i\omega+i\gamma\:\mathrm{sgn}%
(\omega)-\zeta m_{j}}\quad.\label{23}%
\end{equation}
Here it is obvious that the two projection schemes are identical because a
shift of the Matsubara frequencies by $\mu^{\text{ppv}}=-\frac{i\pi}{2\beta}$
becomes irrelevant in the limit $T\rightarrow0$, provided that the Green's
function, or equivalently the fermion spectral-function, is regular at
$\omega=0$. Eq. (\ref{23}) leads to the following self-consistent equations
for the N\'{e}el- and Collinear magnetizations,
\begin{figure}[t]
\begin{center}
\includegraphics*[scale=0.8]{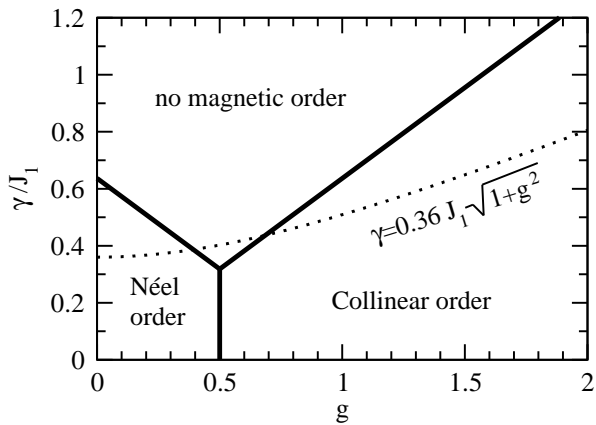}
\end{center}
\caption{Phase diagram in the $\gamma$-$g$-plane. The dotted line shows the
$g$-dependence of $\gamma$ according to Eq. (\ref{22}) for $\tilde{\gamma
}=0.36$.}%
\label{fig3}%
\end{figure}
\begin{subequations}
\begin{align}
&  M_{\text{N\'{e}el}}=\frac{1}{\pi}\arctan\left(  \frac{2M_{\text{N\'{e}el}%
}(J_{1}-J_{2})}{\gamma}\right)  \quad,\label{24a}\\
&  M_{\text{Col}}=\frac{1}{\pi}\arctan\left(  \frac{2M_{\text{Col}}J_{2}%
}{\gamma}\right)  \quad.\label{24b}%
\end{align}
\end{subequations}
The solutions of these equations are shown in Fig. \ref{fig2} for different
parameters $\tilde{\gamma}$. The case $\tilde{\gamma}=0$ represents the
Hartree approximation from Sec. \ref{sec3}. An increase of $\tilde{\gamma}$
reduces the magnetizations, especially in the region of high frustration. In
particular, for small $\tilde{\gamma}$ there is still a direct transition
between the two types of order at $g=\frac{1}{2}$, while for sufficiently
large $\tilde{\gamma}$ a non-magnetic phase emerges. It appears that a
broadening of the pseudofermion levels captures much of the effect of
frustration expected to reduce or destroy magnetic order. In contrast to the
simple Mean-Field theory one now finds second order phase transitions and 
a Mean-Field critical exponent $\beta=\frac{1}{2}$ of the
magnetization. From the self-consistent equations a phase diagram in the
$\gamma$-$g$-plane can be drawn, see Fig. \ref{fig3}. It shows only a narrow
parameter range for $\gamma$ where the theory provides meaningful values for
the phase boundaries. For that reason it will be difficult to determine the
damping parameter in approximative schemes. For example $\tilde{\gamma}=0.36$
leads to transitions at $g_{\text{c1}}\approx0.39$, $g_{\text{c2}}\approx0.69$
and also a realistic value for the magnetization at $g=0$, i.e.,
$M_{\text{N\'{e}el}}\approx0.35$. This value of the width parameter $\tilde{\gamma}$ will be
used in the following section to study the properties of the non-magnetic
phase.

\subsection{Random Phase approximation}{\label{subsec42}} 

\begin{figure}[t]
\begin{center}
$\settowidth{\graphwid}{\includegraphics[scale=0.7,clip=true]{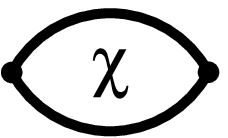}}
\parbox[c]{\graphwid}{\includegraphics[scale=0.7,clip=true]{rpa1t.eps}}=
\settowidth{\graphwid}{\includegraphics[scale=0.7,clip=true]{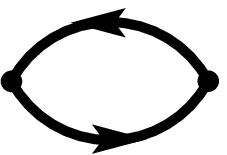}}
\parbox[c]{\graphwid}{\includegraphics[scale=0.7,clip=true]{rpa2t.eps}}+
\settowidth{\graphwid}{\includegraphics[scale=0.7,clip=true]{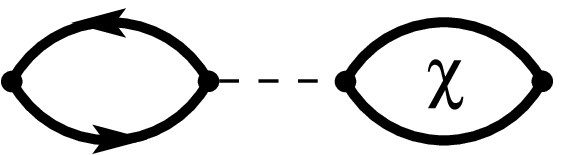}}
\parbox[c]{\graphwid}{\includegraphics[scale=0.7,clip=true]{rpa3t.eps}}$%
\newline$\Pi=
\settowidth{\graphwid}{\includegraphics[scale=0.7,clip=true]{rpa2t.eps}}
\parbox[c]{\graphwid}{\includegraphics[scale=0.7,clip=true]{rpa2t.eps}}$
\end{center}
\caption{Self-consistent RPA equation for the susceptibility $\chi$ in
diagrammatic representation. $\Pi$ denotes a single bubble.}%
\label{fig4}%
\end{figure}

In this section we calculate the spin susceptibility in the
paramagnetic phase within RPA using the Green's function introduced in the
last section. Fig. \ref{fig4} displays the approximation in diagrammatic form.
Since the RPA scheme can be obtained from the Hartree approximation by taking
\begin{figure}[t]
\begin{center}
\includegraphics*[scale=0.8]{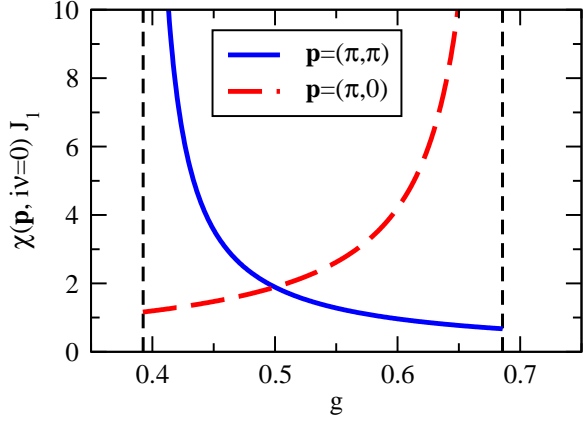}
\end{center}
\caption{Static susceptibility for wave vectors $(\pi,\pi)$ and $(\pi,0)$
and a damping parameter $\tilde{\gamma}=0.36$. The dashed lines visualize the
phase boundaries (color online).}%
\label{fig5}%
\end{figure}
the derivative with respect to the self-consistent field, the quantum phase 
transitions are located at the same point in both approaches. The conserving
approximation scheme in the sense of Baym and Kadanoff \cite{baym1,baym2} is
an essential aspect here because spin conservation is an important constraint on the dynamics of the physical system; in addition the auxiliary particle constraint which allows only one particle per site requires a conserved particle number. In the non-magnetic phase, the equations in Fig.
\ref{fig4} are translation-invariant and can be transformed into momentum
space. The static susceptibility $\chi(\mathbf{p},i\nu=0)$ then has the form
\begin{equation}
\chi(\mathbf{p},i\nu=0)=\frac{1}{\left(  \Pi(i\nu=0)\right)  ^{-1}%
+J(\mathbf{p})}\quad.\label{25}%
\end{equation}
Here $J(\mathbf{p})$ is the Fourier-transform of the interaction $J_{ij}\equiv J_{i-j}$
which is given by
\begin{equation}
J(\mathbf{p})=2J_{1}[\cos(p_{x})+\cos(p_{y})]+4J_{2}\cos(p_{x})\cos(p_{y})\;.\label{26}
\end{equation}
Inserting the propagator from Eq. (\ref{21}) into a single bubble $\Pi
(i\nu=0)$ this quantity is found as
\begin{eqnarray}
&&\Pi(i\nu=0)\equiv\Pi^{zz}(i\nu=0)\notag\\&=&-\frac{1}{4}\frac{1}{2\pi}\int \textrm{d}w \left(\frac{1}{i\omega+i\gamma\:\mathrm{sgn}(\omega)}\right)^{2}\mathrm{Tr}\left[(\sigma^{z})^{2}\right]=\frac{1}{2\pi\gamma}\;.\notag\\ \label{27}
\end{eqnarray}
The susceptibility in Eq. (\ref{25}) together with Eqs. (\ref{26}), (\ref{27})
and (\ref{22}) is evaluated for $\mathbf{p}=(\pi,\pi)$ and $\mathbf{p}%
=(\pi,0)$, the relevant wave vectors in the case of N\'{e}el- and Collinear
\begin{figure}[t]
\begin{center}
\includegraphics*[scale=0.8]{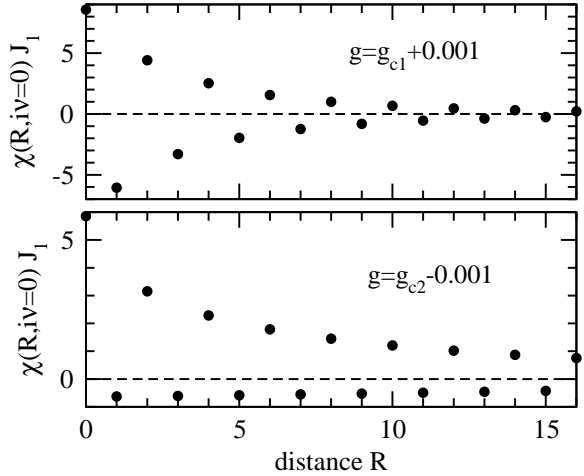}
\end{center}
\caption{Static correlation function $\chi(R,i\nu=0)$ for distances
$R=|\mathbf{R}|$ along a lattice direction. d is measured in units of a
lattice spacing. Again $\tilde{\gamma}=0.36$ is used. In the upper panel $g$
is slightly above the critical point $g_{\mathrm{c1}}$ and in the lower panel
slightly below $g_{\mathrm{c2}}$.}%
\label{fig6}%
\end{figure}
order, respectively. The results are shown in Fig. \ref{fig5}. As expected for
continuous phase transitions, the susceptibility with wave vector $\mathbf{p}%
=(\pi,\pi)$ ($\mathbf{p}=(0,\pi)$) diverges in the limit $g_{c}\rightarrow$
$g_{\mathrm{c1}}+0$ ($g_{c}\rightarrow g_{\mathrm{c2}}-0$).

Finally, we discuss the static correlation function $\chi(\mathbf{R,}i\nu=0)$,
which is obtained by transforming the susceptibility from Eq. (\ref{25}) into
real space,
\begin{equation}
\chi(\mathbf{R,}i\nu=0)=\frac{1}{(2\pi)^{2}}\mathop{\int}_{-\pi}^{\pi
}\hspace*{-1pt}\mathrm{d}p_{x}\hspace*{-1pt}\mathop{\int}_{-\pi}^{\pi}\hspace*{-1pt}\mathrm{d}p_{y}\frac{e^{i\mathbf{pR}%
}}{[\Pi(i\nu=0)]^{-1}+J(\mathbf{p})}.\label{28}%
\end{equation}
Evaluating Eq. (\ref{28}) numerically with $\tilde{\gamma}=0.36$ for distances
$R$ along the vertical or horizontal lattice direction, $\mathbf{R}_{\mu
}=R\mathbf{e}_{\mu}$, $\mu=x,y$, leads to the 
behavior shown in Fig. \ref{fig6}. For $g$ slightly above the
lower critical value $g_{\mathrm{c1}}$ (upper panel) the signature of the
N\'{e}el phase is clearly seen. The correlation function forms a staggered
pattern and the envelopes for positive and negative values only differ by a
sign. At large enough distances $R$ the data points are well fitted by an
exponential decay while at small distances the decrease is faster. Inside the
paramagnetic phase the envelopes are no longer symmetric around $\chi(R)=0$. 
For $g$ slightly below the upper critical point
$g_{\mathrm{c2}}$ (lower panel) the correlation function still exhibits a
staggered sign but the correlation between spins with an odd distance seems to
vanish on approaching the critical point. Again, for large $R$ an exponential
function can be fitted and the correlation length is identical for even and
odd distances. The asymmetry of the two envelopes can be understood by the
fact that for Collinear fluctuations, two degenerate patterns exist, the
alignment of spins along rows and along columns. Thus near the upper critical
point correlations are a superposition of both,
\begin{figure}[t]
\begin{center}
\includegraphics*[scale=0.8]{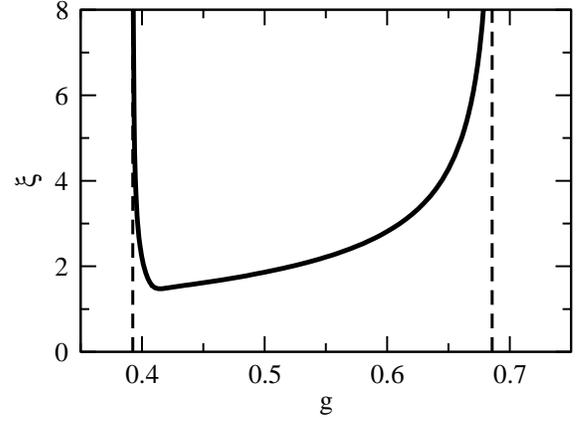}
\end{center}
\caption{Static correlation length $\xi$ in units of a lattice spacing versus
$g$. The spectral width is $\tilde{\gamma}=0.36$.}%
\label{fig7}%
\end{figure}
\begin{equation}
\chi(R)=(-1)^{R}a_{1}e^{-\frac{R}{\xi}}+a_{2}e^{-\frac{R}{\xi}}\label{29}%
\end{equation}
with (almost) identical weights $a_{1}=a_{2}$. Obviously this suppresses
correlations for odd distances. Here $\xi$ denotes the correlation length.
Away from the upper critical point N\'{e}el-like fluctuations emerge and we
have $a_{1}>a_{2}$. Eventually at the lower critical point $a_{2}$ vanishes.

The correlation length $\xi$  is plotted in Fig. \ref{fig7}. The data indicate
divergences at the phase boundaries but get rather small in the vicinity of
$g=0.4$, i.e., down to $\xi\approx1.5$. Remarkably, the smallest values for
the correlation length are not reached at $g=0.5$ where one would classically
expect the strongest frustration.

The phenomenological theory presented suggests that a broadening of the
fermions' spectral function controls the phase diagram and the behavior of
many physical quantities like the magnetization, the susceptibility and the spatial 
correlation function. However, a statement about the nature of the
paramagnetic phase (columnar dimer or plaquette order) can not be made. Also
critical behavior beyond Mean-Field is not accessible. Qualitatively correct
results are obtained by tuning the width $\tilde{\gamma}$. However, this
phenomenological parameter is not calculated within the theory and there is no
simple way to calculate it. Unfortunately, summing up diagrammatic
contributions of the Green's function to gain reasonable values for the width
is a difficult task\cite{jan1,jan2}, because the results strongly depend on
the choice of the diagrams. For example, in the approximation of taking the
self-energy to second order one gets a spectral function of the form
\begin{equation}
\rho(\omega)=\frac{1}{2}(\delta(\omega-\Delta)+\delta(\omega+\Delta))\quad.
\label{30}%
\end{equation}
Then in the whole parameter range $g\geq0$ the gap $\Delta$ turns out to be so
large that magnetic order is destroyed. On the other hand, from a completely
self-consistent calculation of the second order self-energy, one finds that
the spectral function is too narrow to allow for a para\-magnetic phase.

Note that there is no justification for a perturbative treatment in finite
order. Instead diagram classes up to infinite order have to be summed. To
approach this problem in a more systematic way we will now use the Functional
Renormalization Group method (FRG).

\section{Functional Renormalization Group method}{\label{sec5}} 

The FRG method allows to sum up infinite classes of
contributions in perturbation theory in a systematic way.
So far, this method 
has been used to describe the weak coupling regime, e.g. of the Hubbard or
Anderson models. Our model does not have a small coupling constant, since
there is no kinetic energy term in a spin Hamiltonian. We nonetheless employ
FRG in the usual way of neglecting higher order (three particle and higher
order) correlation functions. As it turns out that this is not sufficient, we
add higher order correlations in the form of self-energy corrections. Within
FRG, one starts with the high energy sector, where Green's functions and
coupling functions are known, and successively adds lower energy
contributions. As a first step we define the cutoff procedure to be used by
the following zero temperature bare Green's function,
\begin{equation}
G^{0\Lambda}(i\omega)=\Theta(|\omega|-\Lambda)G^{0}(i\omega)=\frac
{\Theta(|\omega|-\Lambda)}{i\omega+\mu}\quad.\label{31}%
\end{equation}
In this cutoff dependent propagator all modes with $|\omega|<\Lambda$ are
projected out. For the rest of the paper we apply the average auxiliary
fermion projection scheme with $\mu=0$ , as it is exact at zero temperature.
However, it is not too difficult to implement the exact projection scheme\cite{popov},
which increases the numerical effort by roughly a factor of eight. In the
one-particle irreducible (1PI) version of FRG\cite{frg1,frg2,frg3,frg4}
employed here, $G^{0\Lambda}(i\omega)$ is inserted into the generating
functional of the 1PI vertex functions\cite{negele}in place of $G^{0}(i\omega)$. 
Taking the derivative with respect to $\Lambda$, an exact, infinite
hierarchy of coupled differential equations for the vertex functions is
obtained. To be more precise, the flow of the one-particle vertex, the
self-energy $\Sigma$ , depends on $\Sigma$ and the two-particle vertex
$\Gamma$. In turn, the flow of $\Gamma$ depends on $\Sigma$, $\Gamma$ and the
three-particle vertex $\Gamma_{3}$, and so on. At the end of the flow at
$\Lambda=0$ when the theory is cutoff-free, the exact vertex functions are
obtained\cite{frg3,frg4}. However, in explicit calculations one can only deal
with a finite set of equations and hence a truncation scheme has to be
applied. Usually, by applying a weak coupling approximation, the
three-particle vertex $\Gamma_{3}$ and higher vertices are neglected,
resulting in a closed set of equations for $\Sigma$ and $\Gamma$. This scheme
\begin{figure}[t]
\begin{center}
$\frac{d}{d\Lambda}\;\showgraph{scale=0.29}{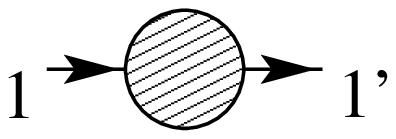}\;=\;-\;\showgraph{scale=0.29}{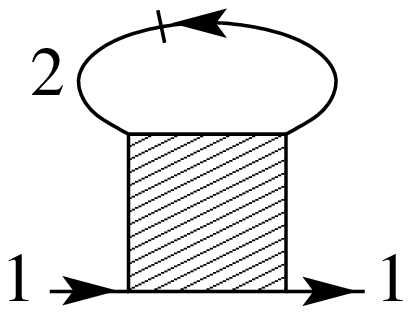}$\\[0.5cm]
$\frac{d}{d\Lambda}\showgraph{scale=0.29}{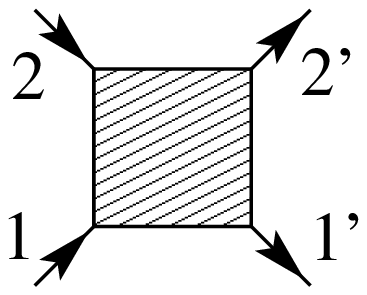}\hspace*{-0.5pt}\textrm{=}\hspace*{-0.5pt} \showgraph{scale=0.29}{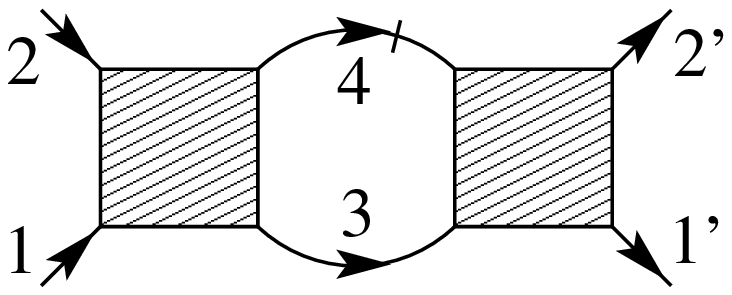}\hspace*{-0.5pt}\textrm{-}\hspace*{-0.5pt}\showgraph{scale=0.29}{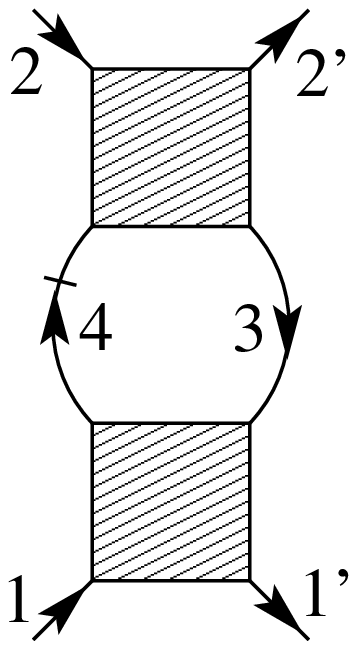}
\hspace*{-0.5pt}\textrm{-}\hspace*{-0.5pt}\showgraph{scale=0.29}{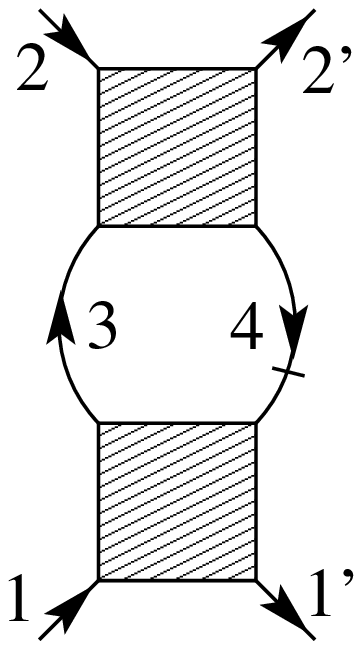}
\hspace*{-1pt}\textrm{+}\hspace*{-1pt}\showgraph{scale=0.29}{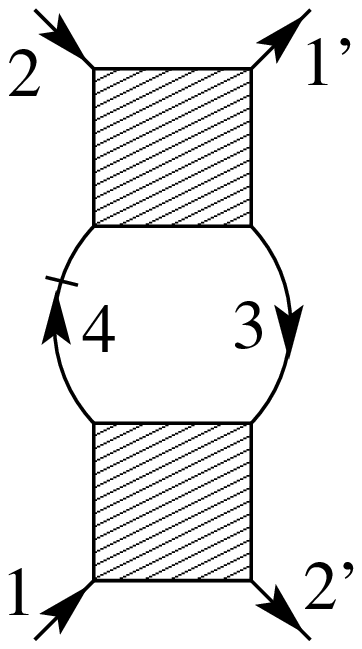}\hspace*{-1pt}\textrm{+}\hspace*{-1pt}\showgraph{scale=0.29}{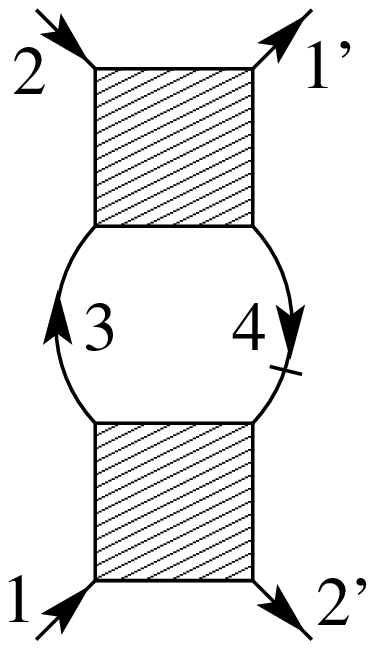}$
\end{center}
\caption{\label{fig8}FRG equations for the self-energy and for the two-particle vertex. The line with an arrow is the full Green's function $G^{\Lambda}(i\omega)$ (see Eq. (\ref{34})) and the line with an arrow and a slash is the single-scale propagator $S^{\Lambda}(i\omega)$ (see Eq. (\ref{35})).}
\end{figure}
will be applied in Secs. \ref{subsec51} and \ref{subsec52}, while in Secs.
\ref{subsec53} and \ref{subsec54} we make use of an improved truncation
scheme\cite{katanin}, which takes into account contributions of the three-particle type. For the conventional truncation scheme the equations for the
self-energy $\Sigma$ and the two-particle vertex $\Gamma$ are depicted in Fig.
\ref{fig8} . In explicit form, these equations read
\begin{equation}
\frac{d}{d\Lambda}\Sigma^{\Lambda}(1)=-\frac{1}{2\pi}\sum_{2}\Gamma^{\Lambda
}(1,2;1,2)S^{\Lambda}(\omega_{2})\label{32}%
\end{equation}%
\begin{eqnarray}
\frac{d}{d\Lambda}\Gamma^{\Lambda}(1',2';1,2)\hspace*{-3pt}&=\hspace*{-4pt}&\frac{1}{2\pi}\sum_{3,4}\left[\Gamma^{\Lambda}(1',2';3,4)\Gamma^{\Lambda}(3,4;1,2)\right.\notag\\
&&-\Gamma^{\Lambda}(1',4;1,3)\Gamma^{\Lambda}(3,2';4,2)\hspace*{-3pt}-\hspace*{-3pt}(3\hspace*{-3pt}\leftrightarrow\hspace*{-3pt}4)\notag\\
&&\left.\hspace*{-1pt}+\Gamma^{\Lambda}(2',4;1,3)\Gamma^{\Lambda}(3,1';4,2)\hspace*{-3pt}+\hspace*{-3pt}(3\hspace*{-3pt}\leftrightarrow\hspace*{-3pt}4)\right]\notag\\
&&G^{\Lambda}(\omega_{3})S^{\Lambda}(\omega_{4})\label{33}
\end{eqnarray} 
Here the numbers are shorthand notations for the frequency, the site index and
the spin index, that is $1=\{\omega_{1},i_{1},\alpha_{1}\}$, and $\sum_{1}$
stands for an integral over $\omega_{1}$ and sums over $i_{1}$ and $\alpha
_{1}$. The full propagator $G^{\Lambda}(i\omega)$ reads \
\begin{equation}
G^{\Lambda}(i\omega)=\frac{\Theta(|\omega|-\Lambda)}{i\omega-\Sigma^{\Lambda
}(i\omega)}\quad,\label{34}%
\end{equation}
and the so-called single-scale propagator is defined by
\begin{equation}
S^{\Lambda}(i\omega)=\left(  G^{\Lambda}(i\omega)\right)  ^{2}\frac
{d}{d\Lambda}\left(  G^{0\Lambda}(i\omega)\right)  ^{-1}=\frac{\delta
(|\omega|-\Lambda)}{i\omega-\Sigma^{\Lambda}(i\omega)}\quad.\label{35}%
\end{equation}
For the last expression in this equation a relationship\cite{morris} for the product 
of $\Theta$-functions and $\delta$-functions has been used.
Note that $G^{\Lambda}$ and $S^{\Lambda}$ are local and translation invariant
in real space and proportional to the unit matrix in spin space. Thus the
propagators in Fig. \ref{fig8} and Eqs. (\ref{32}) and (\ref{33}) carry only
one composite index.

Next we specify the initial conditions for the flow equations at
$\Lambda=\infty$. In this limit, the free propagator vanishes identically.
Thus only one-particle potentials for the self-energy and bare interactions
for the two-particle vertex remain. In the following we confine ourselves to the 
nonmagnetic phase. We defer consideration of the flow of the magnetic order
parameter to later work. Accordingly the free Green's function (\ref{31}) does not contain a
one-particle field $m$ that breaks rotational symmetry as in Eqs. (\ref{8})
and (\ref{10}). Note that although within this scheme the magnetic phases are not
accessible, magnetic instabilities may be detected as divergences in the
susceptibilities. We have a vanishing self-energy for $\Lambda=\infty$,
\begin{equation}
\Sigma^{\Lambda=\infty}(i\omega)\equiv0\quad.\label{36}%
\end{equation}
In this limit, the two-particle vertex is given by the bare interactions in
antisymmetrized form,
\begin{eqnarray}
\Gamma^{\Lambda=\infty}(1',2';1,2)=\hspace*{-4pt}&&\hspace*{-4pt}J_{i_{1}i_{2}}\frac{1}{2}\sigma^{\mu}_{\alpha_{1'}\alpha_{1}}\frac{1}{2}\sigma^{\mu}_{\alpha_{2'}\alpha_{2}}
\delta_{i_{1'}i_{1}}\delta_{i_{2'}i_{2}}\notag\\
&-&\hspace*{-4pt}J_{i_{1}i_{2}}\frac{1}{2}\sigma^{\mu}_{\alpha_{1'}\alpha_{2}}\frac{1}{2}\sigma^{\mu}_{\alpha_{2'}\alpha_{1}}\delta_{i_{1'}i_{2}}\delta_{i_{2'}i_{1}}\quad.\notag\\ \label{37} 
\end{eqnarray}
Here the factors $\frac{1}{2}\sigma_{\alpha\beta}^{\mu}$ originate from the
bare vertices and the Kronecker-$\delta$ ensures that there is no fermion
hopping on the lattice. Since the rotational invariance of the initial conditions 
is conserved during the flow, the two-particle vertex at finite $\Lambda$ is 
parametrized by spin-interaction terms 
$\propto\sigma_{\alpha\beta}^{\mu}\sigma_{\gamma\delta}^{\mu}$ and density-interaction 
terms $\propto\delta_{\alpha\beta}\delta_{\gamma\delta}$. Since the propagators 
are local, the site index of an ingoing leg has to be identical to the site index of 
the corresponding outgoing leg, which results in a total dependence on only two 
sites, i.e.,  $i_{1}$ and $i_{2}$. To be more precise, translation invariance further 
reduces the site dependence only to the separation $|i_{1}-i_{2}|$. Taking into 
account the antisymmetry in all variables the two-particle vertex can now be 
represented as
\begin{eqnarray}
\Gamma^{\Lambda}(1',2';1,2)
=\big{[}\hspace*{-4pt}&&\hspace*{-4pt}\Gamma^{\Lambda}_{\textrm{s}\;i_{1}i_{2}}
(\omega_{1'},\omega_{2'};\omega_{1},\omega_{2})\sigma_{\alpha_{1'}\alpha_{1}}^{\mu}
\sigma_{\alpha_{2'}\alpha_{2}}^{\mu}\notag\\
&+&\hspace*{-4pt}\Gamma^{\Lambda}_{\textrm{d}\;i_{1}i_{2}}(\omega_{1'},\omega_{2'};
\omega_{1},\omega_{2})\delta_{\alpha_{1'}\alpha_{1}}\delta_{\alpha_{2'}\alpha_{2}}\hspace*{4pt}\big{]}\notag\\
&&\hspace*{-18.5pt}\delta_{i_{1'}i_{1}}\delta_{i_{2'}i_{2}}\notag\\
-\hspace*{3pt}\big{[}\hspace*{-4pt}&&\hspace*{-4pt}\Gamma^{\Lambda}_{\textrm{s}\;i_{1}i_{2}}
(\omega_{1'},\omega_{2'};\omega_{2},\omega_{1})\sigma_{\alpha_{1'}\alpha_{2}}^{\mu}
\sigma_{\alpha_{2'}\alpha_{1}}^{\mu}\notag\\
&+&\hspace*{-4pt}\Gamma^{\Lambda}_{\textrm{d}\;i_{1}i_{2}}(\omega_{1'},\omega_{2'};\omega_{2},
\omega_{1})\delta_{\alpha_{1'}\alpha_{2}}\delta_{\alpha_{2'}\alpha_{1}}\hspace*{4pt}\big{]}\notag\\
&&\hspace*{-18.5pt}\delta_{i_{1'}i_{2}}\delta_{i_{2'}i_{1}} \quad.\label{38}
\end{eqnarray}
The indices s/d correspond to spin and density interactions. Note that energy conservation is implied, i.e., 
$\omega_{1'}+\omega_{2'}=\omega_{1}+\omega_{2}$. As another consequence of rotational invariance, 
the self-energy is an odd function in the frequency, as already pointed out after Eq. (\ref{20}). In analogy 
to Eq. (\ref{20}) we write
\begin{equation}
\Sigma^{\Lambda}(i\omega)=-i\gamma^{\Lambda}(\omega)\quad.\label{39} 
\end{equation}
Inserting Eqs. (\ref{34}), (\ref{35}), (\ref{38}) and (\ref{39}) into Eqs. (\ref{32}) and (\ref{33}) the flow 
equations for $\gamma$, $\Gamma_{\textrm{s}}$ and $\Gamma_{\textrm{d}}$ can be calculated.

\subsection{Static FRG}{\label{subsec51}}
Before considering the general case with all frequency dependencies, in this section we briefly discuss 
a static approximation\cite{luttinger,static}. Putting all frequency arguments of the self-energy and vertex 
\begin{figure}[t]
\begin{center}
\includegraphics*[scale=0.8]{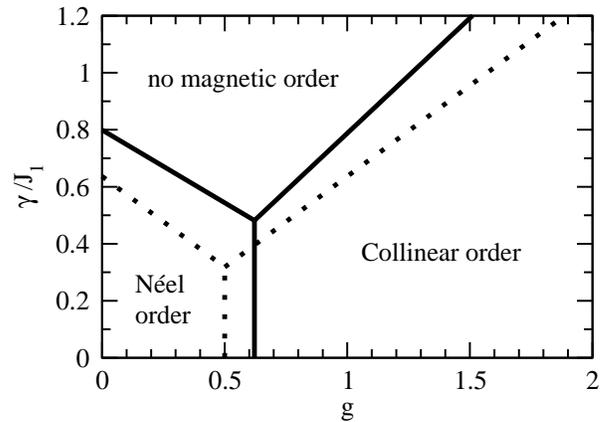}
\end{center}
\caption{\label{fig9}Phase diagram in the $\gamma$-$g$-plane for a static FRG approximation including a 
phenomenological parameter $\gamma$ (full line). The dotted line shows the phase boundaries of the 
phenomenological theory from Fig. \ref{fig3}.}
\end{figure}
functions equal to zero leads, however, to a trivial solution since in that case the self-energy will be identically zero, 
provided it is assumed to be a continuous function of frequency. Therefore, in order to allow for a broadening of the spectrum 
we again assume the discontinuous form $\Sigma^{\Lambda}=-i\gamma^{\Lambda}\:\mathrm{sgn}
(\omega)$. However, inserting this form together with the static two-particle vertex into the first flow equation (\ref{32}) 
leads to a vanishing flow for $\gamma^{\Lambda}$ due to the integration over an odd function on the right side. 
Obviously a static approximation can only be applied to the two-particle vertex, and $\gamma^{\Lambda}$ has to be 
considered again as a phenomenological parameter that is independent of $\Lambda$. Using the static version of Eq. (\ref{38}), i.e., 
\begin{eqnarray}
\Gamma^{\Lambda}\hspace*{-1.4pt}(1'\hspace*{-1.4pt},2'\hspace*{-1.4pt};1\hspace*{-1.4pt},2)\hspace*{-0.75pt}
=\hspace*{-7pt}&&\hspace*{-5pt}\big{[}\Gamma^{\Lambda}_{\textrm{s}\;i_{1}i_{2}}\sigma_{\alpha_{1'}\alpha_{1}}^{\mu}
\sigma_{\alpha_{2'}\alpha_{2}}^{\mu}
\hspace*{-3pt}+\hspace*{-1pt}\Gamma^{\Lambda}_{\textrm{d}\;i_{1}i_{2}}\delta_{\alpha_{1'}\alpha_{1}}
\delta_{\alpha_{2'}\alpha_{2}}\big{]}\notag\\
&&\delta_{i_{1'}i_{1}}\delta_{i_{2'}i_{2}}\notag\\
&-&\hspace*{-5pt}\big{[}\Gamma^{\Lambda}_{\textrm{s}\;i_{1}i_{2}}\sigma_{\alpha_{1'}\alpha_{2}}^{\mu}
\sigma_{\alpha_{2'}\alpha_{1}}^{\mu}
\hspace*{-3pt}+\hspace*{-1pt}\Gamma^{\Lambda}_{\textrm{d}\;i_{1}i_{2}}\delta_{\alpha_{1'}\alpha_{2}}
\delta_{\alpha_{2'}\alpha_{1}}\big{]}\notag\\
&&\delta_{i_{1'}i_{2}}\delta_{i_{2'}i_{1}}\label{40}
\end{eqnarray}
and the phenomenological assumption $\Sigma(i\omega)=-i\gamma\:\mathrm{sgn}(\omega)$ one obtains
\begin{subequations}
\begin{eqnarray}
\frac{d}{d\Lambda}\Gamma^{\Lambda}_{\textrm{s}\;i_{1}i_{2}}&\hspace*{-4pt}=&\hspace*{-4pt}\frac{2}
{\pi}\frac{1}{(\Lambda+\gamma)^{2}}\Big{[}\sum_{j}\Gamma^{\Lambda}_{\textrm{s}\;i_{1}j}
\Gamma^{\Lambda}_{\textrm{s}\;j\,i_{2}}-2\left(\Gamma^{\Lambda}_{\textrm{s}\;i_{1}i_{2}}\right)^{2}\notag\\
&&\hspace*{-4pt}+\Gamma^{\Lambda}_{\textrm{s}\;i_{1}i_{2}}\left(\Gamma^{\Lambda}_{\textrm{s}\;i_{1}i_{1}}-
\Gamma^{\Lambda}_{\textrm{d}\;i_{1}i_{1}}\right)\Big{]}\quad, \label{41a} \\
\frac{d}{d\Lambda}\Gamma^{\Lambda}_{\textrm{d}\;i_{1}i_{2}}&\hspace*{-4pt}=&\hspace*{-4pt}\frac{2}{\pi}\frac{1}
{(\Lambda+\gamma)^{2}}\Big{[}\sum_{j}\Gamma^{\Lambda}_{\textrm{d}\;i_{1}j}
\Gamma^{\Lambda}_{\textrm{d}\;j\,i_{2}}\notag\\
&&\hspace*{-4pt}-\Gamma^{\Lambda}_{\textrm{d}\;i_{1}i_{2}}\left(3\Gamma^{\Lambda}_{\textrm{s}\;i_{1}i_{1}}+
\Gamma^{\Lambda}_{\textrm{d}\;i_{1}i_{1}}\right)\Big{]}\quad. \label{41b}
\end{eqnarray}
\end{subequations}
Note that the frequency dependence of $\Sigma$ only affects the internal integration. By comparing 
Eq. (\ref{37}) and Eq. (\ref{40}) the initial conditions for $\Gamma_{\textrm{s}}$ and $\Gamma_{\textrm{d}}$ can be read off,
\begin{subequations}
\begin{eqnarray}
&&\Gamma_{\textrm{s}\;i_{1}i_{2}}^{\Lambda=\infty}=\frac{1}{4}J_{i_{1}i_{2}}\quad,\label{42a}\\
&&\Gamma_{\textrm{d}\;i_{1}i_{2}}^{\Lambda=\infty}=0\quad.\label{42b}
\end{eqnarray}
\end{subequations}
Solving Eq. (\ref{41b}) with the initial condition (\ref{42b}) gives  $\Gamma_{\textrm{d}\;i_{1}i_{2}}^{\Lambda}\equiv0$. 
A finite set of equations for $\Gamma_{\textrm{s}\;i_{1}i_{2}}^{\Lambda}$ is obtained by neglecting all vertices with 
the distance $|i_{1}-i_{2}|$ exceeding a certain cutoff value. The resulting equations are solved numerically. 
\begin{figure}[t]
\begin{center}
$\frac{d}{d\Lambda}\;\showgraph{scale=0.295}{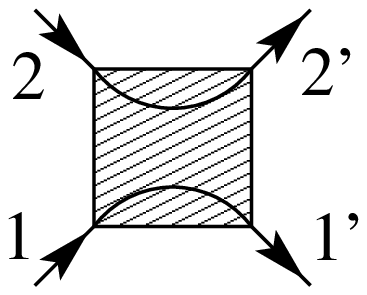}\textrm{=\;-} \showgraph{scale=0.295}{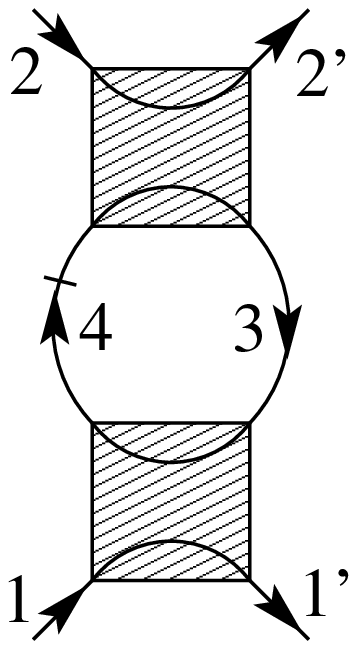}\textrm{-}\showgraph{scale=0.295}{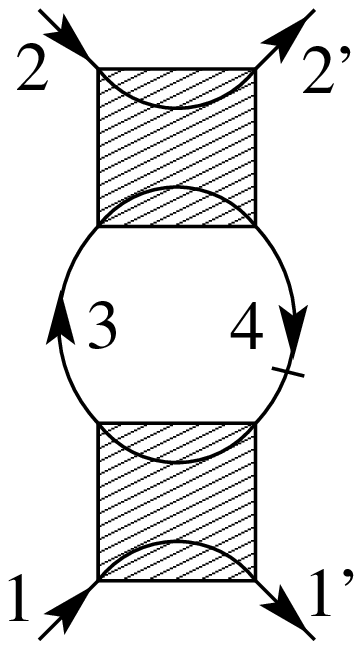}$
\end{center}
\caption{\label{fig10}FRG equation for the RPA scheme. The lines inside the boxes indicate the connections of the external legs.}
\end{figure}
A flow towards finite values for $\Lambda\rightarrow0$ indicates a paramagnetic phase while a diverging flow 
is a sign of a magnetic instability. The type of order can be extracted by transforming $\Gamma_{\textrm{s}
\;i_{1}i_{2}}^{\Lambda}$ into Fourier-space and identifying the fastest momentum component. As a result one 
can draw a phase diagram in the $\gamma$-$g$-plane, see Fig. \ref{fig9}. The figure compares the phase boundaries 
with the results from the phenomenological theory in Sec. \ref{sec4}. Again the boundaries are given by straight lines, 
but the value of the  frustration parameter for which the paramagnetic phase has its largest 
extent moved from $g=0.5$ to $g\approx0.62$. Interestingly, the phase diagram from Sec. \ref{sec4} can be reproduced within FRG. 
Consider the flow equation depicted in Fig. \ref{fig10}. It only contains the second and third term of the right side of the equation 
in Fig. \ref{fig8}. Furthermore it specifies how the ingoing and outgoing lines are connected. An examination of the terms in 
Fig. \ref{fig8} reveals that the contributions in Fig. \ref{fig10} are the only ones that couple two-particle vertices with different 
spatial separations of the outer legs among each other. This is due to the internal fermion bubbles in Fig. \ref{fig10} which 
can be located on an arbitrary site. For the other terms the vertex on the left side of the flow equation is coupled only to itself 
or to the local vertex. The explicit equation corresponding to Fig. \ref{fig10} reads
\begin{equation}
\frac{d}{d\Lambda}\Gamma_{\textrm{s}\;i_{1}i_{2}}^{\Lambda}=\frac{2}{\pi}\frac{1}{(\Lambda+\gamma)^{2}}\sum_{j}
\Gamma_{\textrm{s}\;i_{1}j}^{\Lambda}\Gamma_{\textrm{s}\;j\,i_{2}}^{\Lambda}\quad,\label{43} 
\end{equation}
again with a phenomenological $\gamma$. It can be shown\cite{kat1} that Eq. (\ref{43}) reproduces the static RPA scheme 
and the phase diagram of Sec. \ref{sec4}. Evidently the terms in Fig. \ref{fig10} are essential to obtain magnetism since they 
are the only ones that are able to describe collective phenomena. On the other hand, the remaining terms in Fig. \ref{fig8} are 
only corrections that do not modify the phase diagram qualitatively. 

\subsection{Dynamic FRG}{\label{subsec52}}
The considerations in Sec. \ref{subsec52} led to the conclusion that a static approximation of the FRG equations does not allow to
calculate the central quantity governing the destruction of long range order: the pseudo fermion spectral width $\gamma$ .  We will 
now treat the FRG equations in its full complexity and consider the dynamics with all frequency dependences as well as the 
back coupling of the self-energy into the two-particle vertex. This will lead to a finite spectral broadening without further assumptions. 
Again we make use of the truncation scheme that omits all vertices higher than the two-particle 
vertex. Inserting Eqs. (\ref{34}), (\ref{35}), (\ref{38}) and (\ref{39}) into Eqs. (\ref{32}) and (\ref{33}), after a lengthy but 
straightforward calculation we end up with the flow equations and initial conditions presented in the Appendix. For 
convenience we write the two-particle vertex as a function of the invariant frequency variables $s$, $t$, $u$,
\begin{equation}
\Gamma^{\Lambda}_{\textrm{s/d}\;i_{1}i_{2}}(\omega_{1'},\omega_{2'};\omega_{1},\omega_{2})\rightarrow 
\Gamma^{\Lambda}_{\textrm{s/d}\;i_{1}i_{2}}(s,t,u)\quad,\label{44}
\end{equation}
defined by $s=\omega_{1}+\omega_{2}$, $t=\omega_{1'}-\omega_{1}$, $u=\omega_{1'}-\omega_{2}$. 
The advantage of this parametrization is that $\Gamma_{\textrm{s}}$ and $\Gamma_{\textrm{d}}$ are both invariant 
under each of the transformations $s\rightarrow -s$, $t\rightarrow -t$, $u\rightarrow -u$, which can be deduced by a careful 
examination of the flow equations. This simplifies the numerics  since only positive $s$, $t$, $u$ have to be considered.

In order to solve these equations numerically the continuous frequencies will be discretized. We use a combination of a 
linear and a logarithmic mesh. Since the two-particle vertex depends on three frequencies the computational effort grows 
with the third power of the number of discrete frequencies. Regarding the truncation in real space the computing time grows with 
the fourth power of the length of the longest two-particle vertex (in two dimensions, counting an internal 
site-summation, see Eqs. (\ref{ap2}) and (\ref{ap3})).

From the numerical solution we obtain physical quantities like the static correlation function $\chi_{ij}(i\nu=0)$ by connecting 
the fermion lines of the two-particle vertex,
\begin{eqnarray}
\chi_{ij}(i\nu=0)\hspace*{-2pt}&=&\hspace*{-2pt}\int_{0}^{\infty}\textrm{d}\tau \left\langle T_{\tau}\left\{S_{i}^{z}(\tau)S_{j}^{z}(0)\right\}\right\rangle\notag\\
&=&\hspace*{-2pt}\delta_{ij}\showgraph{scale=0.35}{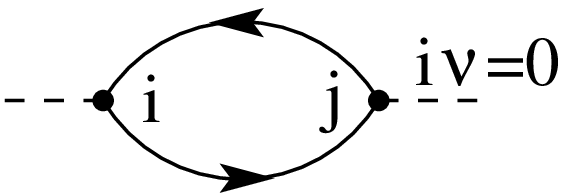}\hspace*{-1pt}+\showgraph{scale=0.35}{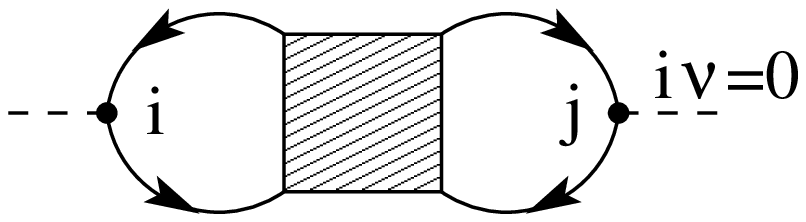}\notag\\ \label{45}
\end{eqnarray}
To calculate this diagram two frequency integrals have to be performed numerically. The physical correlation function is 
recovered in the limit $\Lambda=0$ but we also consider $\chi_{ij}$ at finite $\Lambda$. Transforming $\chi_{ij}^{\Lambda}$ into 
momentum space, we obtain the magnetic susceptibility $\chi^{\Lambda}(\mathbf{p})$. The 
\begin{figure}[t]
\begin{center}
\includegraphics*[scale=0.8]{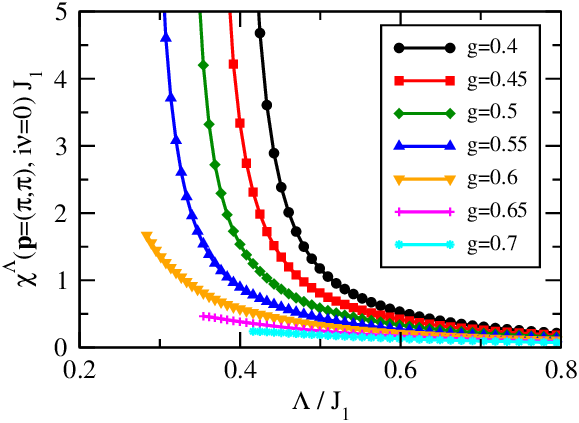}
\end{center}
\caption{\label{fig11}Flow of the static N\'{e}el susceptibility for different frustrations (color online)}
\end{figure}
results are plotted in Fig. \ref{fig11} and Fig. \ref{fig12}. It is seen that during the flow the N\'{e}el susceptibility ($\mathbf{p}=(\pi,\pi)$) exhibits a 
divergence for all $g\lesssim0.55$. On the other hand the Collinear susceptibility appears to diverge for all $g\gtrsim0.55$. 
In particular there is no parameter region without a magnetic instability and with a flow down to $\Lambda=0$. In the present 
approximation the paramagnetic phase is obviously missing. The abrupt stop of the flow of the susceptibilities for $g\geq0.6$ in Fig. \ref{fig11} 
and $g\leq0.55$ in Fig. \ref{fig12} can be traced to the divergence of the respective other susceptibility. For $g$ at the transition, 
i.e., between 0.55 and 0.6, the divergence is clearly indicated at the smallest accessible $\Lambda$.

The absence of a nonmagnetic phase is quite unsatisfactory. Obviously the spectral width comes out too small in this approximation. 
However, an essential improvement is made in the next section where we use a different truncation scheme.

\subsection{Katanin truncation}{\label{subsec53}}
The only approximation that is involved in the FRG scheme described above is the truncation procedure in the hierarchy of 
differential equations. Unfortunately, the simple truncation employed above violates conservation laws, expressed in terms of 
Ward identities. In order to improve the fulfillment of Ward identities Katanin developed a one-particle 
self-consistent version of the two-loop FRG equations\cite{katanin}. The basic modification there is the substitution 
of the single-scale propagator $S^{\Lambda}$, see Eq. (\ref{35}), by the total derivative of $-G^{\Lambda}$ with respect to $\Lambda$,
\begin{equation}
S^{\Lambda}(i\omega)\rightarrow-\frac{d}{d\Lambda}G^{\Lambda}(i\omega)\hspace*{-1pt}=\hspace*{-1pt}S^{\Lambda}(i\omega)-
\left(G^{\Lambda}(i\omega)\right)^{2}\hspace*{-2pt}\frac{d}{d\Lambda}\Sigma^{\Lambda}(i\omega)\;.\label{46} 
\end{equation}
It can be shown\cite{katanin,kat1} that such an approach is equivalent to an RPA + Hartree approximation if only terms of the RPA type (Fig. \ref{fig10}) are kept in the flow equation for the two-particle vertex. 
\begin{figure}[t]
\begin{center}
\includegraphics*[scale=0.8]{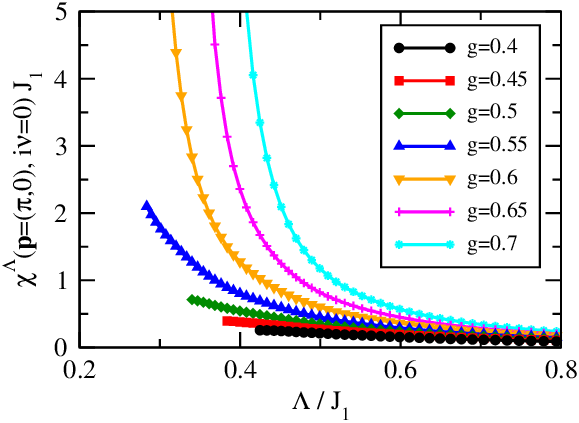}
\end{center}
\caption{\label{fig12}Flow of the static Collinear susceptibility for different frustrations (color online)}
\end{figure}
In this case Ward identities generated by spin conservation are fulfilled exactly. As an application in a different context, for the reduced BCS model of superconductors exact Mean-Field results have been reproduced\cite{kat1}. In particular, in conjunction with a small symmetry-breaking external field this scheme allows to access symmetry broken phases\cite{kat1,kat2,kat3}.  If one keeps the terms additional to RPA on the right hand side of the second flow equation (see Fig. \ref{fig8}, and also Ref. \onlinecite{frg4}), as we do, the exact conservation property is lost, but the remaining symmetry violating terms are generated by overlapping loop diagrams and may be expected to be smaller (see Ref. \onlinecite{katanin}). While on the one hand the Katanin truncation scheme assures that the terms of an RPA+Hartree resummation are correctly included, we find that the non-RPA terms are essential in providing just the right size of a finite auxiliary particle spectral line width. In that sense the non-RPA terms play a crucial role here: they control the pseudofermion damping and therefore the size of the nonmagnetic region in the phase diagram. 

As described in Ref. \onlinecite{kat1} the substitution (\ref{46}) is made in Eq. (\ref{33}) but not in the equation for the self-energy, Eq. (\ref{32}). 
In the present work the above-mentioned small symmetry-breaking field is not applied. This would break the invariance of the 
two-particle vertex under $s\rightarrow-s$, $t\rightarrow-t$, $u\rightarrow-u$ and would generate additional terms in the spin 
parametrization (\ref{38}). Effectively, with the substitution (\ref{46}) also contributions from the three-particle vertex are included. 
In the numerical implementation, in the equations for the two-particle vertex the internal bubble $P_{\textrm{con}}^{\Lambda}$, 
\begin{figure}[t]
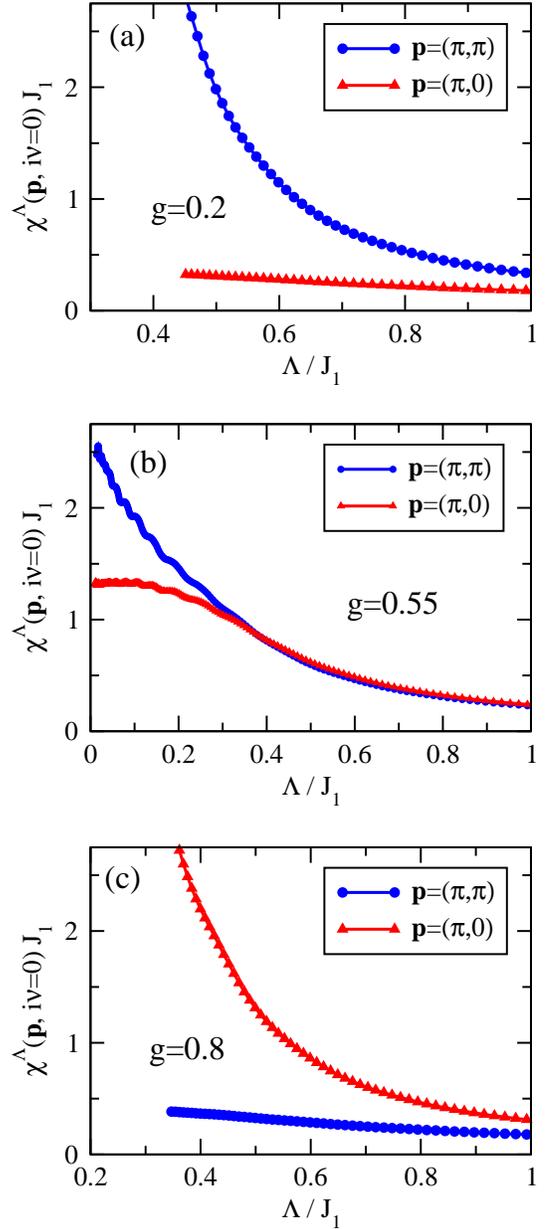

\begin{center}
\includegraphics*[scale=0.8]{susg02.eps}\\[0.4cm]
\includegraphics*[scale=0.8]{susg055.eps}\\[0.4cm]
\includegraphics*[scale=0.8]{susg08.eps}
\end{center}
\caption{\label{fig13}Flow of the static susceptibility for the wave vectors $\mathbf{p}=(\pi,\pi)$ and $\mathbf{p}=(\pi,0)$ and 
different parameters g, (a) $g=0.2$, (b) $g=0.55$, (c) $g=0.8$ (color online).}
\end{figure}
Eq. (\ref{ap5}), is replaced by the modified bubble $P_{\textrm{Kat}}^{\Lambda}$, Eq. (\ref{ap6}). Due to the last term in Eq. (\ref{ap6}) 
which does not contain a $\delta$-function, the internal frequency integration has to be performed explicitly. As a result now the 
computing time grows with the fourth power of the number of discrete frequencies.

Typically we use 64 frequencies and discard all two-particle vertices with a spatial extent larger than 7 lattice spacings in each 
direction. Note that this truncation corresponds to a system with 14$\times$14 sites and periodic boundary conditions because the 
longest bond in such a system extends over 7$\times$7 sites. Exploiting lattice symmetries we end up with approximately 
$2.5\cdot10^{6}$ coupled differential equations. The numerically determined coupling functions and self-energies are inserted into 
Eq. (\ref{45}) to calculate the susceptibilities shown in Fig. \ref{fig13}. In the course of the flow the N\'{e}el susceptibility for $g=0.2$ 
shows a pronounced increase while the Collinear one stays very small, see Fig. \ref{fig13}(a). Obviously at that degree of frustration 
the system is in the N\'{e}el phase. However, we do not observe a real divergence of the susceptibility. When $\Lambda$ 
gets too small the increase of the susceptibility stops and the flow exhibits an unstable and wiggly behavior which we attribute to numerical instabilities. Especially, for small $\Lambda$ the flow is sensitive 
to the discretization of the frequencies. Going to larger system sizes the situation improves, i.e. one can follow the flow to larger 
susceptibilities and finds a steeper increase. Thus, unstable flows at small $\Lambda$ can be identified as finite size effects. 
In the thermodynamic limit and with a sufficient number of discrete frequencies we expect a smooth, diverging solution 
indicating a magnetic instability.

\begin{figure}[t]
\begin{center}
\includegraphics*[scale=0.8]{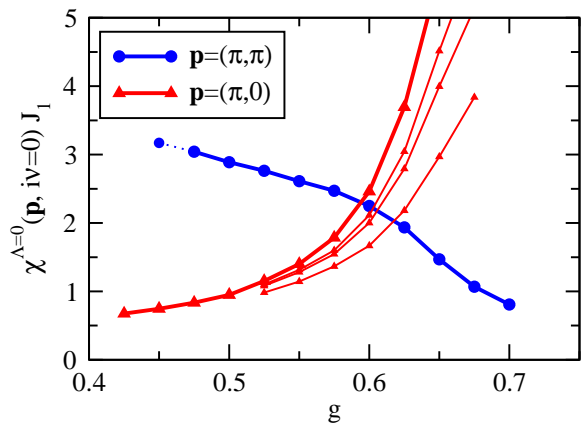}
\end{center}
\caption{\label{fig14}Static susceptibility for the paramagnetic phase in the physical limit $\Lambda=0$. The thick lines are 
obtained by a finite size scaling. The thin lines are the results for a maximal bond length of $7\times7$, $5\times5$ and $3\times3$ 
sites, from top to bottom (color online).}
\end{figure}
At $g=0.55$, see Fig. \ref{fig13}(b), both susceptibilities approach finite values for $\Lambda\rightarrow0$, demonstrating the existence 
of a phase with neither N\'{e}el nor Collinear long range order. Small oscillations are a consequence of the frequency mesh. In 
our numerics the limit $\Lambda=0$ can not be reached exactly because of an insufficient number of discrete frequencies at very low energy scales. 
Typically the flow is stopped at $\Lambda\approx0.01J_{1}$ but can be easily extrapolated to $\Lambda=0$. Finally at $g=0.8$ 
the Collinear phase can be identified. In Fig. \ref{fig13}(c) the behavior is analogous to Fig. \ref{fig13}(a), but showing an 
increasing Collinear susceptibility.

In order to investigate the properties of this phase further, we calculated the susceptibilities at additional 
parameter values. The results in the physical limit $\Lambda=0$ are shown in Fig. \ref{fig14}. Deep inside the paramagnetic phase 
our results are well-converged. With increasing $g$ we observe a decreasing N\'{e}el susceptibility and an increasing Collinear 
susceptibility. The point where N\'{e}el-like fluctuations loose out compared to Collinear fluctuations lies at $g\approx0.6$ in correspondence 
with the results in Sec. \ref{subsec51}, i.e., clearly higher than the classical value $g=0.5$. At the phase boundary to Collinear order, 
which turns out to be in the range $g_{\textrm{c2}}\approx0.66\ldots0.68$, the critical fluctuations require large system sizes, in order to obtain 
well-converged results. Here a finite size scaling (see thin red lines in Fig. \ref{fig14}) considerably enhances the Collinear 
susceptibility and a beginning divergence is visible. The situation is very different near the phase boundary to N\'{e}el order. A divergence 
\begin{figure}[t]
\begin{center}
\includegraphics*[scale=0.8]{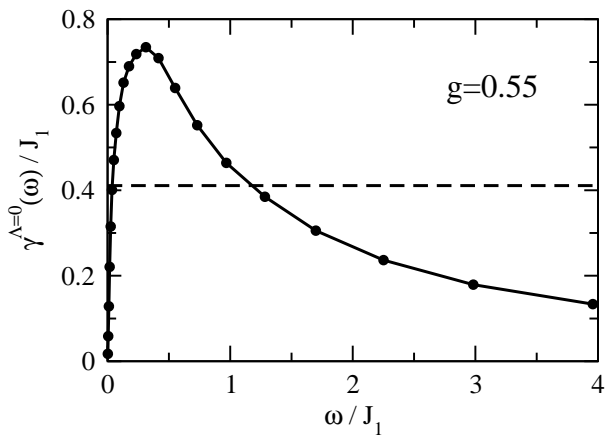}
\end{center}
\caption{\label{fig141}Frequency dependent damping $\gamma^{\Lambda=0}(\omega)$ obtained at the end of the FRG-flow compared to the constant damping with $\tilde{\gamma}=0.36$ used in Sec. \ref{sec4} (dashed line). Depicted is the case $g=0.55$.}
\end{figure}
\begin{figure}[t]
\begin{center}
\includegraphics*[scale=0.6]{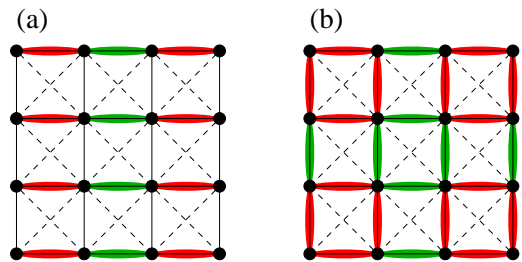}
\end{center}
\caption{\label{fig15}Patterns for (a) columnar dimerization and (b) plaquette dimerization. Red bonds correspond to strengthened and green 
bonds to weakened interactions in the Hamiltonian $H+F_{\textrm{d}}$ or $H+F_{\textrm{p}}$ (color online).}
\end{figure}
of the N\'{e}el susceptibility is not seen and finite size effects play a minor role. Instead, here the flow is highly sensitive to the frequency 
discretization, which causes large oscillations. Therefore it seems to be difficult to access this critical region and to obtain reliable data, 
see the dotted part of the blue line in Fig. \ref{fig14}. Here a denser mesh enhances and smooths the N\'{e}el susceptibility during 
the flow. An estimation of this phase boundary leads to the parameter region $g_{\textrm{c1}}\approx0.4\ldots0.45$.

Only few results on spin susceptibilities are found in the literature\cite{mod6}. To the best of our knowledge so far no data are available 
for static spin susceptibilities. We note that our results on the phase boundaries are in good agreement with previous 
results\cite{mod6,mod12,mod16,dimer2,darradi,isaev}.

So far we notice that the behavior of the system near the two transitions is very different. To draw a conclusion concerning the order 
of the transitions, a closer investigation taking into account a flowing order parameter is necessary.

Although the tendency towards formation of N\'{e}el- and Collinear phases has already been seen in the standard truncation of the 
previous subsection, the inclusion of certain higher order terms turns out to be essential. As shown above, the non-RPA like terms in conjunction with the Katanin truncation indeed lead to a damping $\gamma^{\Lambda}(\omega)$ which is strong enough to generate a non-magnetic phase. The damping which is related to the self-energy via Eq. (\ref{39}) is no physical observable. This quantity is obtained from the first flow equation and can be compared with the frequency independent $\gamma$ used in the phenomenological theory from Sec. \ref{sec4}, see Fig. \ref{fig141}. At $g=0.55$, i.e., in the non-magnetic region we find that $\gamma^{\Lambda=0}(\omega)$ compares quite well with the choice $\tilde{\gamma}=0.36$ in Eq. (\ref{22}) at relevant energy scales $\omega\sim J$. In the region with high frustration, $g\approx0.55$, the static two-particle vertex $\Gamma_{\textrm{s}\;i_{1}i_{2}}^{\Lambda}(s=0,t=0,u=0)$, where $i_{1}$, $i_{2}$ are nearest neighbors and using the initial conditions given by Eq. (\ref{ap7}), comes out as $\Gamma_{\textrm{s}\;i_{1}i_{2}}^{\Lambda=0}(0,0,0)\approx1.8J_{1}$ at the end of the flow.

\subsection{Columnar dimer and plaquette order}{\label{subsec54}}
In this section we discuss the nature of the paramagnetic phase and investigate whether there is still some kind of long 
range order. Possible states currently under discussion are a spin liquid state (which does not break any symmetries) 
and a VBS state. For the VBS two dimerization patterns are of special interest: In a columnar dimer arrangement 
translation invariance along one lattice direction as well as rotation symmetry are broken. For a plaquette 
valence-bond ordering, translation symmetry in both directions is broken while the rotation symmetry is intact.

In order to probe the paramagnetic phase with respect to these states we add a small perturbation field to the Hamiltonian and investigate 
the response to it\cite{dimer1,dimer2,dimer3,dimer4,sirker,darradi,isaev}. In the context of FRG this concept has already been applied in Refs. \onlinecite{kat1,kat2}. The fields can be chosen as
\begin{eqnarray}
&&\hspace*{-12pt}F_{\textrm{d}}=\delta\sum_{i,j}(-1)^{i}\mathbf{S}_{i,j}\mathbf{S}_{i+1,j}\;,\notag\\ 
&&\hspace*{-12pt}F_{\textrm{p}}=\delta\sum_{i,j}\left[(-1)^{i}\mathbf{S}_{i,j}\mathbf{S}_{i+1,j}+(-1)^{j}\mathbf{S}_{i,j}\mathbf{S}_{i,j+1}\right]\;,\label{47} 
\end{eqnarray}
\begin{figure}[t]
\begin{center}
\includegraphics*[scale=0.8]{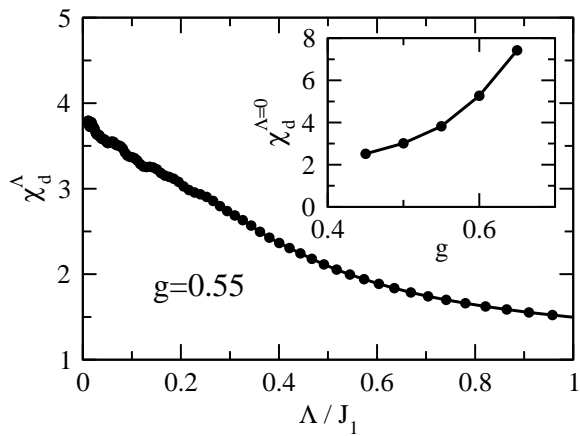}
\end{center}
\caption{\label{fig16}\textbf{Main figure:} Flowing dimer correlation $\chi_{\textrm{d}}^{\Lambda}$ for $g=0.55$. \textbf{Inset:} Dimer 
correlation $\chi^{\Lambda=0}_{\textrm{d}}$ in the limit $\Lambda=0$ versus $g$.}
\end{figure}
for the columnar dimer and plaquette order, respectively. Here $i$, $j$ are components of the position vector and $\delta$ is an energy 
much smaller than $J_{1}$ and $J_{2}$. Note that the expectation values $\langle F_{\textrm{d}}\rangle$ and $\langle F_{\textrm{p}}\rangle$ 
are the order parameters of these states. $F_{\textrm{d}}$ (see Refs. \onlinecite{dimer1,dimer2,dimer3,dimer4,sirker,darradi,isaev}) and 
$F_{\textrm{p}}$ (see Refs. \onlinecite{darradi,isaev}) break the above-mentioned lattice symmetries and generate the two dimerization 
patterns shown in Fig. \ref{fig15}. Possible instabilities should be visible as divergences in the corresponding equal time correlation functions 
$\chi_{\textrm{d/p}}=\frac{d\langle F_{\textrm{d/p}}\rangle}{d\delta}|_{\delta=0}$.

The coupling to these operators is included in the FRG formalism by modifying the initial conditions. The bare interactions in the limit 
$\Lambda\rightarrow\infty$ are slightly strengthened or 
weakened according to the dimerization patterns. Furthermore we have to take into account that due to the broken translation symmetries 
a two-particle vertex is no longer uniquely determined by one lattice vector. 
In the course of the flow we calculate the correlations of strengthened and weakened bonds and its relative difference. We define equal time dimer and plaquette correlation functions by
\begin{equation}
\chi^{\Lambda}_{\textrm{d/p}}=
\frac{J_{1}}{\delta}\frac{\left|\left(\langle\hspace*{-1pt}\langle\mathbf{S}_{i,j},\mathbf{S}_{i+1,j}\rangle\hspace*{-1pt}\rangle^{\Lambda}_{\textrm{d/p}}-
\langle\hspace*{-1pt}\langle\mathbf{S}_{i+1,j},\mathbf{S}_{i+2,j}\rangle\hspace*{-1pt}\rangle^{\Lambda}_{\textrm{d/p}}
\right)\right|}
{\left(\langle\hspace*{-1pt}\langle\mathbf{S}_{i,j},\mathbf{S}_{i+1,j}\rangle\hspace*{-1pt}\rangle^{\Lambda}_{\textrm{d/p}}+
\langle\hspace*{-1pt}\langle\mathbf{S}_{i+1,j},\mathbf{S}_{i+2,j}\rangle\hspace*{-1pt}\rangle^{\Lambda}_{\textrm{d/p}}
\right)}\;\;.\label{48} 
\end{equation}
Here the index d/p indicates that the correlator $\langle\hspace*{-1pt}\langle\ldots\rangle\hspace{-1pt}\rangle$ is calculated with 
the Hamiltonian $H+F_{\textrm{d/p}}$. The factor $\delta$ in 
the denominator eliminates the dependence on the strength of the perturbation such that $\chi_{\textrm{d/p}}^{\Lambda}$ start with the initial 
value $\chi_{\textrm{d/p}}^{\Lambda=\infty}=1$. An increase (decrease) during the flow shows that the system supports (rejects) the 
perturbation. Note that we again apply Katanin's truncation scheme.

The columnar dimer correlation $\chi_{\textrm{d}}^{\Lambda}$ is plotted in Fig. \ref{fig16}. It is seen that this quantity increases 
considerably during the flow. In the limit $\Lambda=0$ the perturbation is enhanced by a factor of $\approx3.8$ but a divergence does 
not occur. We do not exclude an instability for this kind of order which might be masked here due to finite size effects. Remarkably 
we obtain plaquette correlations $\chi_{\textrm{p}}^{\Lambda}$ with the same strength. Apparently the FRG scheme is not able to distinguish between dimer and plaquette correlations. This might be a consequence of the fact that the four-particle vertex from which such susceptibilities can be calculated directly, is not included in our FRG equations. The oscillations during the flow are again a consequence of the frequency mesh.

Thus our results favor a spin liquid with enhanced equal time correlations $\chi_{\textrm{d}}$, $\chi_{\textrm{p}}$ or a VBS with one of the two 
arrangements. Previous papers mainly calculated the columnar dimer and plaquette susceptibilities in the magnetically ordered 
phases rather than in the paramagnetic phase\cite{dimer2,dimer3,sirker,darradi}.

\section{Summary}{\label{sec6}}
The aim of this work is the development of new methods and the calculation of properties for frustrated quantum spin models. 
We focused on the spin-1/2 $J_{1}$-$J_{2}$ Heisenberg antiferromagnet on the square lattice, but our method is generally 
applicable to models of that type. Starting point of our approach is perturbation theory in the exchange couplings, summed 
to infinite order. In order to be able to use standard many-body techniques and to perform diagrammatic expansions, we 
applied the auxiliary-fermion representation of spin operators. To enforce the auxiliary-particle constraint, 
two different projection schemes have been employed: (1) enforcement of the constraint on the average (which, however, 
becomes exact at zero temperature), and (2) exact projection using an imaginary chemical potential\cite{popov} .

In a first exploratory study we use the RPA + Hartree approximation to access the ground state properties. Since the straightforward 
Mean-Field approach is not able to describe the suppression of magnetic order near $g=\frac{J_{2}}{J_{1}}=0.5$, we introduce 
a damping term in the bare pseudofermion Green's function. This phenomenological self-energy accounts for 
scattering processes of the auxiliary fermions which lead to a finite lifetime and a spectral broadening. We show that the damping 
reduces the magnetic order especially in the regime of strong frustration. For sufficiently large damping one finds a paramagnetic 
phase around $g=\frac{J_{2}}{J_{1}}=0.5$, as seen in numerical studies (we term this phase paramagnetic although it may 
possess more complex magnetic correlations)  . Furthermore, using RPA, we calculate the 
magnetic susceptibility and the spin correlations in the non-magnetic phase. We observe critical behavior at the phase 
boundaries, i.e., a divergent susceptibility and correlation length. Within this method the basic properties come out qualitatively 
well, but a microscopic derivation of the pseudofermion damping is beyond the reach of simple diagrammatic resummations. 

A more systematic approach is considered in the main part of the paper: the Functional Renormalization Group method (FRG). 
We employ its formulation in terms of the one-particle irreducible vertex functions and use a sharp frequency cutoff $\Lambda$. This 
method sums up large diagram classes in a systematic way, e.g. the two-particle vertex function in the particle-particle channel 
and in two particle-hole channels and reaches therefore far beyond the Mean-Field theory. Self-energy contributions are taken 
into account on equal footing with the vertex renormalizations. 

First we apply the conventional truncation scheme to the hierarchy of FRG differential equations, neglecting all vertices 
higher than the two-particle vertex. Further imposing the static approximation, we find that the pseudofermion broadening 
is not generated in this way. Adding a phenomenological broadening, the results of the phenomenological theory are recovered, 
which is saying that RPA-like diagrams in the particle-hole channel can be identified as the most important contributions.
Including the frequency dependencies of the self-energy and the two-particle vertex we find magnetic instabilities in the whole parameter 
range. The latter is signalled by an immanent divergence of the susceptibilities at $\mathbf{p}=(\pi,\pi)$ and/or  $\mathbf{p}=(\pi,0)$ in the course 
of the flow towards $\Lambda=0$. It thus turns out that the truncation scheme is insufficient to generate a strong 
pseudofermion damping and a nonmagnetic phase. This can be traced to the violation of Ward identities in this approximation.

An improved approximation including self-energy effects in the single-scale propagator has been suggested by Katanin\cite{katanin}. 
There the single-scale propagator is replaced by the total derivative of the Green's function with respect to $\Lambda$. 
We find that the latter approach, even if only implemented on the one-loop level, reproduces features of the Mean-Field theory 
and fulfills the Ward identities exactly, if the RPA-like contributions in the particle-hole channel are considered only\cite{katanin,kat1}. 
In our calculations we include the additional terms on the right side of the equation for the two-particle vertex in Fig. \ref{fig8}. 
Calculating the susceptibility, we are now able to distinguish between the three phases. In particular we get a convergent flow 
down to $\Lambda=0$ in the region where the paramagnetic phase is expected. The phase boundary between 
N\'{e}el-order and paramagnetic phase is found to be at  $g_{\textrm{c1}}\approx0.4\ldots0.45$ and the
transition from paramagnetism to Collinear order happens at  $g_{\textrm{c2}}\approx0.66\ldots0.68$. Our findings agree 
well with the results obtained by  other methods\cite{mod6,mod12,mod16,dimer2,darradi,isaev}. Approaching the transition 
to Collinear order we observe a smooth divergence of the corresponding susceptibility. On the other hand near the transition 
to N\'{e}el-order a different picture emerges. Here it is difficult to access the critical region because the discretization of the frequencies generates large oscillations in the flow at small $\Lambda$.

Finally we probed the non-magnetic phase with respect to columnar dimerization and plaquette order by investigating the 
flow in the presence of appropriate small perturbative fields. In the limit $\Lambda=0$ the correlations for both types of order are enhanced but a divergence is not found. 
These results indicate either strong dimer and plaquette fluctuations in a spin-liquid phase or a symmetry broken phase with 
dimer or plaquette order.

The work reported here shows that in spite of the fact that quantum spin models are in the strong coupling regime by definition, 
partial resummations of perturbation theory appear to capture the physics of frustrated magnets at least on a qualitative level. 
The resummations, done here in the framework of the Functional Renormalization Group method, account in a controlled and 
systematic way for all two-particle interaction processes, including all couplings between the different channels.  If self-energy 
corrections are added in a consistent way, thereby including certain contributions of the three-particle vertex, it appears to be possible to calculate 
the central quantity of frustrated spin systems in the language of pseudoparticles, the damping of the pseudofermions 
in a controlled way. The systems we consider are in principle infinitely large but the spin correlations are only treated up to some finite length. Hence we do not have to deal with the effects of edges or periodic boundary conditions. However, the range over which our correlations extend typically include more than 200 sites which is much more than the system sizes accessible by exact diagonalization. Furthermore we do not make any assumption on the ground state or perform an expansion around any presumed state. Our starting point of free fermions without dispersion is completely featureless. Therefore our approach is straightforwardly applicable to a variety of models. Some obvious extensions of the present work are (1) consideration of other models such as the spin-1/2 $J_{1}$-$J_{2}$-$J_{3}$ Heisenberg antiferromagnet on the square lattice, or geometrically frustrated models like the triangular or Kagome lattice, (2) generalization to finite temperature, (3) calculation of dynamical spin correlation functions. Work in this direction is in progress.

\begin{acknowledgements}
We thank W. Brenig, R. Thomale, H. Schmidt, M. Salmhofer, and S. Andergassen for stimulating discussions. Financial support by the Deutsche Forschungsgemeinschaft through the Forschergruppe FOR 960 is gratefully acknowledged.
\end{acknowledgements}

\appendix*
\section{FRG equations with full dynamics}\label{dix1}
In this appendix we show the flow equations for $\gamma$, $\Gamma_{\textrm{s}}$ and $\Gamma_{\textrm{d}}$ with all frequency dependences,
\begin{widetext}
\begin{eqnarray}
\frac{d}{d\Lambda}\gamma^{\Lambda}(\omega)=\frac{1}{2\pi}
\Big{[}-2\sum_{j}\big{(}&&\hspace*{-12pt}\Gamma_{\textrm{d}\;ij}^{\Lambda}(\omega+\Lambda,0,\omega-\Lambda)-\Gamma_{\textrm{d}\;ij}^{\Lambda}(\omega-\Lambda,0,\omega+\Lambda)\big{)}\notag\\
+3\big{(}&&\hspace*{-12pt}\Gamma_{\textrm{s}\;i\hspace*{0.5pt}i\hspace*{1pt}}^{\Lambda}(\omega+\Lambda,\omega-\Lambda,0)
-\Gamma_{\textrm{s}\;i\hspace*{0.5pt}i\hspace*{0.5pt}}^{\Lambda}(\omega-\Lambda,\omega+\Lambda,0)\big{)}\notag\\
+&&\hspace*{-12pt}\Gamma_{\textrm{d}\;i\hspace*{0.5pt}i\hspace*{0.5pt}}^{\Lambda}(\omega+\Lambda,\omega-\Lambda,0)
-\Gamma_{\textrm{d}\;i\hspace*{0.5pt}i\hspace*{0.5pt}}^{\Lambda}(\omega-\Lambda,\omega+\Lambda,0)\Big{]}\frac{1}{\Lambda+\gamma^{\Lambda}(\Lambda)}
\quad,\label{ap1} 
\end{eqnarray}
\begin{eqnarray}
\frac{d}{d\Lambda}\Gamma^{\Lambda}_{\textrm{s}\;i_{1}i_{2}}(s,t,u)&=&\frac{1}{2\pi}\int_{-\infty}^{\infty}\textrm{d}\omega'
\Big{[}\hspace*{5pt}\big{(}-2\Gamma^{\Lambda}_{\textrm{s}\;i_{1}i_{2}}(s,-\omega_{2'}-\omega',\omega_{1'}+\omega')\Gamma^{\Lambda}_{\textrm{s}\;i_{1}i_{2}}(s,\omega_{2}+\omega',\omega_{1}+\omega')\notag\\
&&\hspace*{76pt}+\Gamma^{\Lambda}_{\textrm{s}\;i_{1}i_{2}}(s,-\omega_{2'}-\omega',\omega_{1'}+\omega')\Gamma^{\Lambda}_{\textrm{d}\;i_{1}i_{2}}(s,\omega_{2}+\omega',\omega_{1}+\omega')\notag\\
&&\hspace*{76pt}+\Gamma^{\Lambda}_{\textrm{d}\;i_{1}i_{2}}(s,-\omega_{2'}-\omega',\omega_{1'}+\omega')\Gamma^{\Lambda}_{\textrm{s}\;i_{1}i_{2}}(s,\omega_{2}+\omega',\omega_{1}+\omega')
\big{)}\notag \\
&&\hspace*{62pt}\big{(}P^{\Lambda}(\omega',s+\omega')+P^{\Lambda}(s+\omega',\omega')\big{)}\notag\\
&&\hspace*{54pt}+\big{(}2\sum_{j}\Gamma^{\Lambda}_{\textrm{s}\;i_{1}j}\hspace*{2.5pt}(\omega_{1'}+\omega',t,\omega_{1}-\omega')\Gamma^{\Lambda}_{\textrm{s}\;j\,i_{2}}\hspace*{2pt}(\omega_{2}+\omega',t,-\omega_{2'}+\omega')\notag\\
&&\hspace*{81pt}+\Gamma^{\Lambda}_{\textrm{s}\;i_{1}i_{2}}(\omega_{1'}+\omega',t,\omega_{1}-\omega')\Gamma^{\Lambda}_{\textrm{s}\;i_{1}i_{1}}(\omega_{2}+\omega',-\omega_{2'}+\omega',t)\notag\\
&&\hspace*{81pt}-\Gamma^{\Lambda}_{\textrm{s}\;i_{1}i_{2}}(\omega_{1'}+\omega',t,\omega_{1}-\omega')\Gamma^{\Lambda}_{\textrm{d}\;i_{1}i_{1}}(\omega_{2}+\omega',-\omega_{2'}+\omega',t)\notag\\
&&\hspace*{81pt}+\Gamma^{\Lambda}_{\textrm{s}\;i_{1}i_{1}}(\omega_{1'}+\omega',\omega_{1}-\omega',t)\Gamma^{\Lambda}_{\textrm{s}\;i_{1}i_{2}}(\omega_{2}+\omega',t,-\omega_{2'}+\omega')\notag\\
&&\hspace*{81pt}-\Gamma^{\Lambda}_{\textrm{d}\;i_{1}i_{1}}(\omega_{1'}+\omega',\omega_{1}-\omega',t)\Gamma^{\Lambda}_{\textrm{s}\;i_{1}i_{2}}(\omega_{2}+\omega',t,-\omega_{2'}+\omega')
\big{)}\notag\\
&&\hspace*{62pt}\big{(}P^{\Lambda}(\omega',t+\omega')+P^{\Lambda}(t+\omega',\omega')\big{)}\notag\\
&&\hspace*{54pt}-\big{(}2\Gamma^{\Lambda}_{\textrm{s}\;i_{1}i_{2}}(\omega_{2'}-\omega',-\omega_{1}-\omega',u)\Gamma^{\Lambda}_{\textrm{s}\;i_{1}i_{2}}(\omega_{2}-\omega',\omega_{1'}+\omega',u)\notag\\
&&\hspace*{63.5pt}+\Gamma^{\Lambda}_{\textrm{s}\;i_{1}i_{2}}(\omega_{2'}-\omega',-\omega_{1}-\omega',u)\Gamma^{\Lambda}_{\textrm{d}\;i_{1}i_{2}}(\omega_{2}-\omega',\omega_{1'}+\omega',u)\notag\\
&&\hspace*{63.5pt}+\Gamma^{\Lambda}_{\textrm{d}\;i_{1}i_{2}}(\omega_{2'}-\omega',-\omega_{1}-\omega',u)\Gamma^{\Lambda}_{\textrm{s}\;i_{1}i_{2}}(\omega_{2}-\omega',\omega_{1'}+\omega',u)
\big{)}\notag\\
&&\hspace*{62pt}\big{(}P^{\Lambda}(\omega',u+\omega')+P^{\Lambda}(u+\omega',\omega')\big{)}\Big{]}\quad,\label{ap2}
\end{eqnarray}
\begin{eqnarray}
\frac{d}{d\Lambda}\Gamma^{\Lambda}_{\textrm{d}\;i_{1}i_{2}}(s,t,u)&=&\frac{1}{2\pi}\int_{-\infty}^{\infty}\textrm{d}\omega'
\Big{[}\hspace*{5pt}\big{(}3\Gamma^{\Lambda}_{\textrm{s}\;i_{1}i_{2}}(s,-\omega_{2'}-\omega',\omega_{1'}+\omega')\Gamma^{\Lambda}_{\textrm{s}\;i_{1}i_{2}}(s,\omega_{2}+\omega',\omega_{1}+\omega')\notag\\
&&\hspace*{63.5pt}+\Gamma^{\Lambda}_{\textrm{d}\;i_{1}i_{2}}(s,-\omega_{2'}-\omega',\omega_{1'}+\omega')\Gamma^{\Lambda}_{\textrm{d}\;i_{1}i_{2}}(s,\omega_{2}+\omega',\omega_{1}+\omega')
\big{)}\notag \\
&&\hspace*{62pt}\big{(}P^{\Lambda}(\omega',s+\omega')+P^{\Lambda}(s+\omega',\omega')\big{)}\notag\\
&&\hspace*{54pt}+\big{(}2\sum_{j}\Gamma^{\Lambda}_{\textrm{d}\;i_{1}j}\hspace*{2.5pt}(\omega_{1'}+\omega',t,\omega_{1}-\omega')\Gamma^{\Lambda}_{\textrm{d}\;j\,i_{2}}\hspace*{2pt}(\omega_{2}+\omega',t,-\omega_{2'}+\omega')\notag\\
&&\hspace*{76pt}-3\Gamma^{\Lambda}_{\textrm{d}\;i_{1}i_{2}}(\omega_{1'}+\omega',t,\omega_{1}-\omega')\Gamma^{\Lambda}_{\textrm{s}\;i_{1}i_{1}}(\omega_{2}+\omega',-\omega_{2'}+\omega',t)\notag\\
&&\hspace*{81pt}-\Gamma^{\Lambda}_{\textrm{d}\;i_{1}i_{2}}(\omega_{1'}+\omega',t,\omega_{1}-\omega')\Gamma^{\Lambda}_{\textrm{d}\;i_{1}i_{1}}(\omega_{2}+\omega',-\omega_{2'}+\omega',t)\notag\\
&&\hspace*{76pt}-3\Gamma^{\Lambda}_{\textrm{s}\;i_{1}i_{1}}(\omega_{1'}+\omega',\omega_{1}-\omega',t)\Gamma^{\Lambda}_{\textrm{d}\;i_{1}i_{2}}(\omega_{2}+\omega',t,-\omega_{2'}+\omega')\notag\\
&&\hspace*{81pt}-\Gamma^{\Lambda}_{\textrm{d}\;i_{1}i_{1}}(\omega_{1'}+\omega',\omega_{1}-\omega',t)\Gamma^{\Lambda}_{\textrm{d}\;i_{1}i_{2}}(\omega_{2}+\omega',t,-\omega_{2'}+\omega')
\big{)}\notag\\
&&\hspace*{62pt}\big{(}P^{\Lambda}(\omega',t+\omega')+P^{\Lambda}(t+\omega',\omega')\big{)}\notag\\
&&\hspace*{54pt}-\big{(}3\Gamma^{\Lambda}_{\textrm{s}\;i_{1}i_{2}}(\omega_{2'}-\omega',-\omega_{1}-\omega',u)\Gamma^{\Lambda}_{\textrm{s}\;i_{1}i_{2}}(\omega_{2}-\omega',\omega_{1'}+\omega',u)\notag\\
&&\hspace*{63.5pt}+\Gamma^{\Lambda}_{\textrm{d}\;i_{1}i_{2}}(\omega_{2'}-\omega',-\omega_{1}-\omega',u)\Gamma^{\Lambda}_{\textrm{d}\;i_{1}i_{2}}(\omega_{2}-\omega',\omega_{1'}+\omega',u)
\big{)}\notag\\
&&\hspace*{62pt}\big{(}P^{\Lambda}(\omega',u+\omega')+P^{\Lambda}(u+\omega',\omega')\big{)}\Big{]}\quad.\label{ap3}
\end{eqnarray}
\end{widetext}
Note that the frequency parametrization of Eq. (\ref{44}) is used for $\Gamma_{\textrm{s}}$ and $\Gamma_{\textrm{d}}$. The frequencies $\omega_{1'}$, $\omega_{2'}$, $\omega_{1}$, $\omega_{2}$ on the right side stand for the inverse transformations
\begin{eqnarray}
\omega_{1'}=\frac{1}{2}(s+t+u)&,&\omega_{2'}=\frac{1}{2}(s-t-u)\;\;,\notag\\
\omega_{1}=\frac{1}{2}(s-t+u)&,&\omega_{2}=\frac{1}{2}(s+t-u)\;\;.\label{ap4} 
\end{eqnarray}
$P^{\Lambda}(\omega_{1},\omega_{2})$ denotes a bubble of $S^{\Lambda}$ and $G^{\Lambda}$. For the conventional truncation as discussed in Sec. \ref{subsec52} one gets
\begin{equation}
P^{\Lambda}(\omega_{1},\omega_{2})\rightarrow P^{\Lambda}_{\textrm{con}}(\omega_{1},\omega_{2})=\frac{\delta(|\omega_{1}|-\Lambda)}{\omega_{1}+\gamma^{\Lambda}(\omega_{1})} \frac{\Theta(|\omega_{2}|-\Lambda)}{\omega_{2}+\gamma^{\Lambda}(\omega_{2})}\;.\label{ap5}
\end{equation}
In this scheme the internal integration $\int\textrm{d}\omega'\cdots$ simplifies to $\sum_{\omega'=\pm \Lambda}\cdots\;$. For the Katanin truncation considered in Sec. \ref{subsec53} we get a more complicated expression,
\begin{eqnarray}
&&\hspace*{-10pt}P^{\Lambda}(\omega_{1},\omega_{2})\rightarrow P^{\Lambda}_{\textrm{Kat}}(\omega_{1},\omega_{2})=\frac{\delta(|\omega_{1}|-\Lambda)}{\omega_{1}+\gamma^{\Lambda}(\omega_{1})} \frac{\Theta(|\omega_{2}|-\Lambda)}{\omega_{2}+\gamma^{\Lambda}(\omega_{2})}\notag\\
&&\hspace*{-10pt}+\left(\frac{d}{d\Lambda}\gamma^{\Lambda}(\omega_{1})\right)
\frac{\Theta(|\omega_{1}|-\Lambda)}{(\omega_{1}+\gamma^{\Lambda}(\omega_{1}))^{2}} \frac{\Theta(|\omega_{2}|-\Lambda)}{\omega_{2}+\gamma^{\Lambda}(\omega_{2})}\quad.\label{ap6} 
\end{eqnarray}
In both cases $P^{\Lambda}(\omega_{1},\omega_{2})$ is an odd function separately in $\omega_{1}$ and $\omega_{2}$. Finally from the comparison between Eq. (\ref{37}) and Eq. (\ref{38}) we get the following initial conditions,
\begin{eqnarray}
&\gamma^{\Lambda=\infty}(\omega)=0\quad,&\notag\\
&\Gamma_{\textrm{s}\;i_{1}i_{2}}^{\Lambda=\infty}(s,t,u)=\frac{1}{4}J_{i_{1}i_{2}}\;\;,\;\;\Gamma_{\textrm{d}\;i_{1}i_{2}}^{\Lambda=\infty}(s,t,u)=0\;\;.&\label{ap7} 
\end{eqnarray}

\bibliography{paper_revised}

\end{document}